\documentclass{aa}  

\usepackage{graphicx}
\usepackage{txfonts}
\newcommand{\sG}{{\,$\sigma$\,Gem}}

\begin{document}

\title{Observing the changing surface structures of the active K giant $\sigma$\,Gem with SONG\thanks{Based on observations made with the Hertzsprung SONG telescope on the Spanish Observatorio del Teide on the island of Tenerife and the Tennessee State University T3 0.4-m Automated Photometric Telescope at Fairborn Observatory in Arizona.}}

   \author{H. Korhonen\inst{1}
          \and
          R.M. Roettenbacher\inst{2}
          \and
          S. Gu\inst{3,4,5}
          \and
          F. Grundahl\inst{6}
          \and
          M. F. Andersen\inst{6}
          \and
          G.W. Henry\inst{7}
          \and
          J. Jessen-Hansen\inst{6}
          \and
          V. Antoci\inst{8,6}
          \and
          P.L. Pall{\'e}\inst{9,10}
          }

   \institute{
     European Southern Observatory (ESO), Alonso de C{\'o}rdova 3107, Vitacura, Santiago, Chile\\
     \email{heidi.korhonen@eso.org}
     \and
     Yale Center for Astronomy and Astrophysics, Department of Physics, Yale University, New Haven, CT 06520, USA
     \and
     Yunnan Observatories, Chinese Academy of Sciences, Kunming 650216, Yunnan Province, China
     \and
     Key Laboratory for the Structure and Evolution of Celestial Objects, Chinese Academy of Sciences, Kunming 650216, China
     \and
     School of Astronomy and Space Science, University of Chinese Academy of Sciences, Beijing 101408, China
     \and
     Stellar Astrophysics Centre, Department of Physics and Astronomy, Aarhus University, Ny Munkegade 120, DK-8000 Aarhus C, Denmark
     \and
     Center of Excellence in Information Systems, Tennessee State University, Nashville, TN 37209, USA
     \and
     DTU Space, National Space Institute, Technical University of Denmark, Elektrovej 328, DK-2800 Kgs. Lyngby, Denmark
     \and
     Instituto de Astrof{\'i}sica de Canarias, E-38205 La Laguna, Tenerife, Spain
     \and
     Universidad de La Laguna, Departamento de Astrof{\'i}sica, E-38206 La Laguna, Tenerife, Spain
   }
   \date{Received September 15, 1996; accepted March 16, 1997}

  \abstract
   {}
   {We aim to study the spot evolution and differential rotation in the magnetically active cool K-type giant star \sG\ from broadband photometry and continuous spectroscopic observations that span 150 nights.}
   {We use high-resolution, high signal-to-noise ratio spectra obtained with the Hertzsprung SONG telescope to reconstruct surface (photospheric) temperature maps with Doppler imaging techniques. The 303 observations span 150 nights and allow for a detailed analysis of the spot evolution and surface differential rotation. The Doppler imaging results are compared to simultaneous broadband photometry from the Tennessee State University T3 0.4 m Automated Photometric Telescope. The activity from the stellar chromosphere, which is higher in the stellar atmosphere, is also studied using SONG observations of Balmer H$\alpha$ line profiles and correlated with the photospheric activity. 
 }
   {The temperature maps obtained during eight consecutive stellar rotations show mainly high-latitude or polar spots, with the main spot concentrations above latitude 45$^{\circ}$. The spots concentrate around phase 0.25 near the beginning of our observations and around phase 0.75 towards the end. The photometric observations confirm a small jump in spot phases that occurred in February 2016. The cross-correlation of the temperature maps reveals rather strong solar-like differential rotation, giving a relative surface differential rotation coefficient of $\alpha=0.10\pm0.02$. There is a weak correlation between the locations of starspots and enhanced emission in the chromosphere at some epochs.}
   {}


   \maketitle
%

\section{Introduction}

It is widely accepted that the global behaviour of solar and stellar magnetic fields can be explained by dynamo action, which is due to interaction between magnetic fields and fluid motions. The Sun and Sun-like stars are thought to harbour a dynamo in which the poloidal field is created from the toroidal field by helical convection, and the toroidal field is obtained by shearing the already existing poloidal field via differential rotation \citep[for a review see, e.g.][]{2010LRSP....7....3C,2017LRSP...14....4B}. Insight into these dynamo processes can be obtained by studying the cool photospheric spots created by the magnetic fields. 

In the solar case, tracking the movement of sunspots has been used to study solar differential rotation, for example by \cite{1986A&A...155...87B} and \cite{2017A&A...606A..72P}. Doing the same in the stellar case is more challenging due to a lack of spatial and temporal resolution in the available observations \citep[for different stellar techniques see, e.g.][]{2012IAUS..286..268K,2013AN....334...89C}. Detailed mapping of the stellar surfaces is only possible with Doppler imaging techniques \citep[see, e.g.][]{1983PASP...95..565V,2008MNRAS.385..708S,2016A&A...594A..29B,2018A&A...620A.162J} and long-baseline optical interferometry \citep{2016Natur.533..217R}, the latter of which has only recently become available. However, these stellar techniques typically offer snapshots of the stellar surface -- often months or even years apart. Robotic facilities allow observations to be obtained at a higher cadence, which enables the study of the stellar surface from one rotation to the next to track the surface features more accurately.

In recent years, studies investigating stellar surface features from continuous spectroscopic observations obtained with robotic facilities have been published. For example, \cite{2015A&A...578A.101K} studied the spot evolution and surface differential rotation on the RS\,CVn-type binary XX\,Tri, using six years of data from the robotic STELLA facility \citep{2004AN....325..527S}. \cite{2015A&A...578A.101K} detected weak, solar-like surface differential rotation operating on XX\,Tri and also determined from the spot-decay rate a turbulent diffusivity of $\eta T = 6.3 \pm 0.5 \times 10^{14}$\,cm$^2$s$^{-1}$. Other authors have also used continuous observations from robotic facilities to study stellar surface differential rotation and spot evolution, including \cite{2016A&A...592A.117H}, \cite{2016A&A...596A..53K}, and \cite{2017A&A...597A.101F}.

Here, we analyse continuous spectroscopic observations of the RS\,CVn-type binary \object{$\sigma$\,Gem} (HD 62044). This system has been classified as a single-lined spectroscopic binary by \cite{1955ApJ...121..118H}. The secondary was detected with a primary-to-secondary H-band flux ratio of $270\pm70$ via long-baseline optical interferometry \citep{2015ApJ...807...23R}. The primary is a very well-studied, magnetically active K giant. It shows all the normal characteristics of RS\,CVn-type binaries: frequent flaring \citep[see, e.g.][]{2006ApJ...638L..37B,2013ApJ...768..135H}, chromospheric activity \citep[see, e.g.][]{1988ApJS...68..803B,2000A&AS..146..103M}, and brightness variations caused by large starspots \citep[see, e.g.][]{1989A&A...218..192O,1995ApJS...97..513H,2014A&A...562A.107K}. These cool surface spots have also been mapped with Doppler imaging techniques \citep[e.g.][]{1993ApJ...410..777H,2001A&A...373..199K}. Additionally, \cite{2017ApJ...849..120R} studied the detailed surface structures on \sG\ using light curve, Doppler, and interferometric imaging methods.

However, while \sG\ has been widely studied, only results from two campaigns to continuously study the spot evolution on its surface have been published. The first continuous observations were done by \cite{2001A&A...373..199K}, who studied the evolution of surface spots for 3.6 stellar rotations in 1996--1997. Unfortunately, they were not able to clearly determine spot evolution and differential rotation from these observations. They attributed this to the possible masking effect of short-term spot evolution. A re-analysis of their data by \citep{2007A&A...474..165K} resulted in the detection of weak anti-solar differential rotation, where the polar regions rotate faster than the equator \citep{2007A&A...474..165K}. The second continuous campaign to study \sG\ was published by \cite{2015A&A...573A..98K}. The authors re-analysed the 1996--1997 observations as well as data from 2006--2007 obtained during approximately five consecutive rotations. Anti-solar differential rotation was also detected in these data.

In this paper, we analyse the longest continuous spectroscopic data set of \sG\ published to date. This data set reveals clear evolution in stellar magnetic activity on \sG. In Sect.\,\ref{sect:obs}, we describe the observations, data processing, and our data analysis methods. In Sect.\,\ref{sect:results}, we present the temperature maps obtained using Doppler imaging and also study the chromospheric activity using H$\alpha$ spectral line profiles. In Sect.\,\ref{sect:disc}, we discuss our results and set them in the context of earlier studies.  We summarise our main conclusions in Sect.\,\ref{sect:conc}.

\section{Observations and methods}
\label{sect:obs}

\subsection{Photometric observations}

\begin{figure}
  \centering
  \includegraphics[width=8cm]{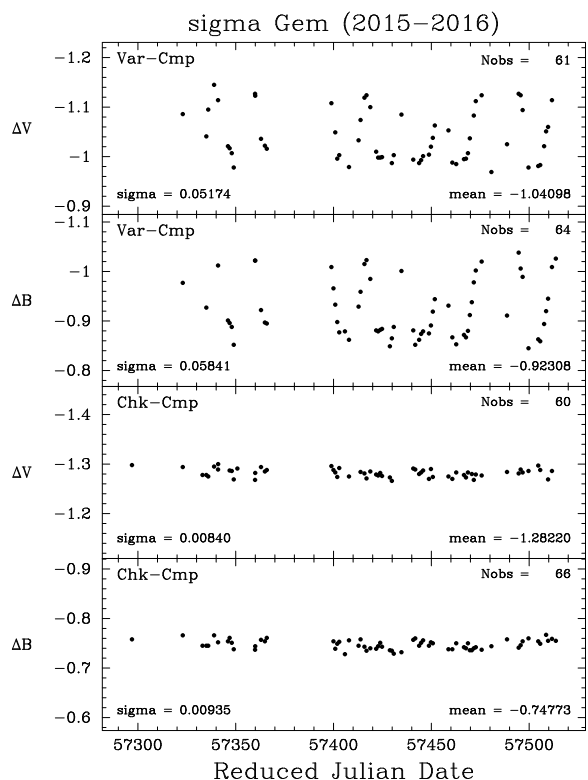}
  \caption{Photometric observations of \sG\ acquired with the Tennessee State University T3 0.4 m Automated Photometric Telescope between October 2015 and May 2016.}
  \label{photometry}
\end{figure}

Our target, \sG, was observed between 27 October 2015 and 5 May 2016 with the Tennessee State University T3 0.4~m Automated Photometric Telescope at Fairborn Observatory in Arizona \citep{1999PASP..111..845H}. The observations consist of Johnson {\it B} and {\it V} differential magnitudes, which are defined as the variable star (\sG ) minus the comparison star (Var-Cmp).  To ensure the observed variability is that of the variable star, differential magnitudes of a check star minus the comparison star (Chk-Cmp) were also obtained.  The comparison star was HD~60318 while the check star was HD~60522.  The (Chk-Cmp) observations do not show any trends or periodicities; their standard deviations are 0.00935~mag in {\it B} and 0.00840~mag in {\it V}, consistent with the long-term precision of the APT. All (Var-Cmp) and (Chk-Cmp) measurements are shown in Fig.\,\ref{photometry}.

From interferometric, photometric and spectroscopic observations, \cite{2015ApJ...807...23R} determined that the primary of \sG\ is starting to fill its Roche-lobe, and its shape is slightly ellipsoidal with a major-to-minor axis ratio of $1.02\pm0.03$. Before analysing the photometric observations for the presence of starspots, we first removed the ellipsoidal variations using the procedure of \cite{2017ApJ...849..120R}. The light curve is dominated by spots, and its shape is barely altered by the removal of the ellipsoidal variability. The original differential magnitudes and corresponding differential magnitudes with the ellipsoidal variations removed are listed in Table\,\ref{obs_log_photom}. 

\subsection{Spectroscopic observations}

During the 2015--2016 observing season, \sG\ was observed by the Stellar Observations Network Group \citep[SONG,][]{2008JPhCS.118a2041G} node at the Teide Observatory on Tenerife, Spain. The observations with the Hertzsprung SONG telescope began on 4 November 2015 and ended 1 April 2016. Throughout these 150 nights, spectra of \sG\ were acquired almost nightly; often multiple spectra were obtained in a single night. The SONG programme yielded 303 spectra covering the full time period. Observations where the optimal extraction failed, namely with signal-to-noise ratio (SNR) less than 50 in the main spectral region of interest, were discarded. This yielded 265 spectra suitable for Doppler imaging. The full observing log stating the time of the observations and the SNR obtained is given in Table\,\ref{obs_log_spec}.

The Hertzsprung SONG telescope is a 1-metre telescope equipped with a high-resolution spectrograph \citep{2017ApJ...836..142G} and a Lucky Imaging Camera \citep{2015A&A...574A..54S}. The facility is operated in a fully automated mode \citep{2019PASP..131d5003F}. We used the spectrograph with slit \#5, giving a resolving power ($\lambda/\Delta\lambda$) of 77,000. The spectra cover wavelengths from 4400\,{\AA} to 6900\,{\AA} in 51 orders, with some small gaps redwards of 5300\,{\AA}. Each individual spectrum of \sG\ had an exposure time of 180 seconds. This typically resulted in a SNR per pixel of $\sim$200 measured in a continuum region around at 6450\,{\AA}. However, due to weather, the fluctuations in SNR were large and the exact values for individual exposures are given in Table\,\ref{obs_log_spec}. 

\subsection{Reduction and further processing of the spectroscopic data}

The spectroscopic data were reduced by the SONG pipeline, \emph{Songwriter}. The pipeline uses the standard calibration frames (bias, flat fields and ThAr lamp spectra) that are obtained daily. The reduction pipeline is written in Python, except for the optimal extraction for which the method of \citet{2002A&A...385.1095P} re-implemented by \cite{2014PASP..126..170R} is used. All the \sG\ data are freely available in the SONG Data Archive, SODA\footnote{https://soda.phys.au.dk/}.

The SONG data cover almost eight consecutive rotations of \sG. Due to bad weather, there are a few gaps towards the end of our 150 night observing period. To obtain good phase coverage for Doppler imaging, the data were divided into seven subsets, each covering between 0.84 and 1.62 stellar rotations. More details on the subsets are given in Table\,\ref{subsets}, including the dates covered for each subset, the number of spectra obtained, and the rotational cycles spanned by each subset. 

\begin{table}
  \caption{Division of the SONG data into eight subsets.}
  \label{subsets}
  \centering          
  \begin{tabular}{l l l l}
    \hline\hline          
    Subset & Observing dates & No of & Rotational \\
           &                 & Spectra   & Cycle\tablefootmark{a} \\ 
    \hline
    set01 & 2015-11-04 -- 2015-11-22 & 20 & 0.12 -- 1.04 \\
    set02 & 2015-11-24 -- 2015-12-12 & 24 & 1.14 -- 2.07 \\
    set03 & 2015-12-23 -- 2015-12-31 & 28 & 2.12 -- 3.03 \\
    set04 & 2016-01-04 -- 2016-01-21 & 78(17)\tablefootmark{b} & 3.23 -- 4.14 \\
    set05 & 2016-01-23 -- 2016-02-09 & 67(12)\tablefootmark{b} & 4.24 -- 5.10 \\
    set06 & 2016-02-09 -- 2016-02-26 & 30 & 5.10 -- 5.94 \\
    set07 & 2016-03-01 -- 2016-04-01 & 26 & 6.14 -- 7.76 \\
\hline                                             
  \end{tabular}
  \tablefoot{
    \tablefoottext{a}{The phases for the rotational cycle are calculated with the ephemeris given in Table\,\ref{sigmaGem_param} and set01 set as the cycle zero.}
    \tablefoottext{b}{The first number gives the total number of spectra obtained this time period, and the second number in parentheses is the number of higher SNR spectra obtained by combining the observations obtained in groups of sequential spectra.  Details of the combined spectra are found in Table\,\ref{obs_log_spec}}
}
\end{table}

In set04 and set05 there are several rotational phases during which many spectra were obtained closely in time, and some of those individual spectra have SNR of 75 or less. Therefore, for these two subsets, the data obtained immediately after each other were combined into one higher SNR spectrum. The longest time period within which the data were combined was 105 minutes. This corresponds to about 0.4\% of the stellar rotation and does not introduce any noticeable phase smearing. This resulted in 17 spectra for set04 and 12 for set05. We also note that for improving the phase coverage, on one occasion the same data point was used in two different sets, namely phase 0.105 at MJD 57427.8 in set05 and set06. 

Good phase coverage is an important factor in producing reliable Doppler images. Optimally, one would need at least ten data points uniformly spread over the rotation period \citep[see, e.g.][]{2000A&AS..147..151R}. Towards the end of the SONG observing period the observations become sparser, at times affecting the phase coverage. Therefore, the spectra from the last 1.6 stellar rotations are used to obtain only one temperature map. We are confident that no significant changes have happened in the spot configuration during this time period (see Appendix\,\ref{set07} for more details) and we note that often 1.5 stellar rotations or more are used in Doppler imaging even for relatively slow rotators \citep[see, e.g.][]{2015A&A...574A..31S,2015A&A...573A..98K}.

The \emph{Songwriter} pipeline does not merge the spectral orders, and additionally the orders redwards of 5300\,{\AA} do not overlap. Order 44 covers most of the spectral lines typically used for Doppler imaging, including three \ion{Fe}{I} lines (at 6419\,{\AA}, 6421\,{\AA}, and 6430\,{\AA}), and one \ion{Ca}{I} line (at 6439\,{\AA}). We used these four spectral lines simultaneously for the temperature mapping.

The continuum correction for the whole Doppler imaging region was done by fitting a second degree polynomial to the regions known to be in the continuum, based on synthetic spectra and our previous high SNR, high-resolution spectra \citep[see ][]{2017ApJ...849..120R}. The continuum level for individual phases was further fine-tuned during the Doppler imaging process.

In our analysis of the chromospheric activity, we used the  H$\alpha$ line. This spectral region has numerous small water lines arising from the Earth's atmosphere, so-called telluric lines. At times, depending on the exact orbital phase, these lines were also superimposed on the H$\alpha$ line profile. To improve the analysis of the variations in the H$\alpha$ line profile we removed the telluric lines using the \emph{Molecfit} tool \citep{2015A&A...576A..77S}. The amount of precipitable water vapour was varied to account for the changing amount of water in the Earth's atmosphere. Visual inspection of the H$\alpha$ line profiles showed that the water lines were removed to a high degree.

\subsection{Stellar parameters}

In Doppler imaging the observations are compared to a grid of synthetic model line profiles. Precise stellar parameters are needed to calculate them accurately.

Our target, \sG\, is often used for Doppler imaging, and it also has very accurately determined orbital parameters. Here, we calculate the phases using the ephemeris and orbit determined by \cite{2015ApJ...807...23R}. That determination utilises both direct interferometric detection of the secondary and radial velocities of both stars. We used the stellar parameters determined by \cite{2017ApJ...849..120R} for our Doppler imaging. The only exception is the rotational velocity. The best-fit to the SONG data was obtained with a somewhat larger rotational velocity, $v\sin i$ of 26.2 km\,s$^{-1}$, than the 24.8 km\,s$^{-1}$ obtained by \cite{2017ApJ...849..120R}. This $v\sin i$ value is still lower than 27.5$\pm$1 km\,s$^{-1}$ determined by \cite{2001A&A...373..199K}. We also note that the interferometric imaging gives an inclination of 107.73$^{\circ}$ \citep{2017ApJ...849..120R}, but as the Doppler imaging cannot distinguish the orientation of the star on the sky, we used an inclination less than 90$^{\circ}$. All stellar parameters adopted for Doppler imaging are given in Table\,\ref{sigmaGem_param}. 

\begin{table}
  \caption{Stellar parameters of \sG\ adopted for Doppler imaging. These parameters are from \cite{2017ApJ...849..120R}, with the exception of $v\sin i$, which is from this work.}
  \label{sigmaGem_param}
  \centering          
  \begin{tabular}{l l}
    \hline\hline          
    Parameter      &  Value \\
    \hline                                           
    Effective Temperature, $T_{eff}$ [K] & 4530     \\
    Inclination, $i$ [degrees]          & 72 \\
    Metallicity, Fe/H                   & 0.0 \\
    Surface gravity, $\log g$           & 2.5 \\
    Rotational velocity, $v\sin i$ [km\,s$^{-1}$] & 26.2 \\
    Orbital period, $P_{orb}$ [days]     & 19.6030 \\
    Time of nodal passage, $T_{0}$ [MJD] & 53583.61 \\
    Microtubulence [km\,s$^{-1}$]        & 0.8 \\
    Macrotubulence [km\,s$^{-1}$]        & 2.0 \\
\hline                                             
\end{tabular}
\end{table}

\subsection{Doppler imaging process and line profile modelling}

The surface temperature maps were obtained using Doppler imaging techniques \citep[see, e.g.][]{1983PASP...95..565V}. Here, we used the code \emph{INVERS7PD}, which was originally developed by \cite{1990A&A...230..363P} and later modified by \cite{2001A&A...374..171H}. Doppler imaging is an ill-posed inverse problem, and \emph{INVERS7PD} uses Tikhonov regularisation for solving it.  

The grid of synthetic local line profiles were calculated in the same way as described by \cite{2017ApJ...849..120R}. The grid consists of twenty limb angles, nine temperatures (3500--5500\,K with a 250\,K step in between), and wavelengths ranging from 6408.5\,{\AA} to 6441.0\,{\AA} with a wavelength step of 0.01\,{\AA}. In the inversion, the stellar surface was divided into a grid of 40 latitudes and 80 longitudes. Before inversion, the local line profiles were convolved with a Gaussian instrumental profile and a radial-tangential macroturbulence velocity. The same code and models were used for calculating the photometric output, but with a sparser wavelength grid ranging from 3600\,{\AA} to 7350\,{\AA} and a step size of 50\,{\AA}.

\begin{figure*}
  \centering
  \includegraphics[width=18cm]{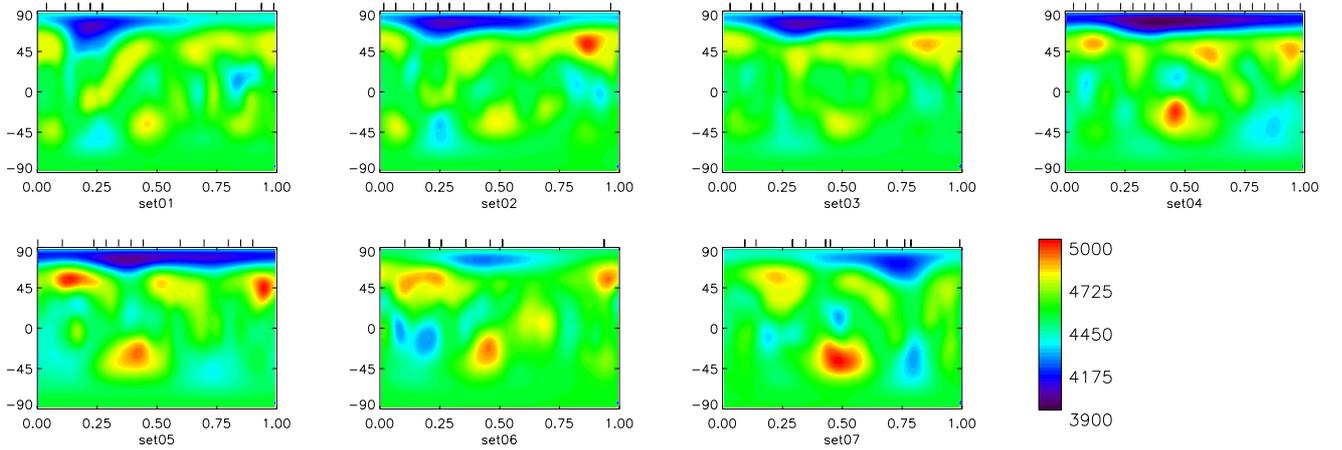}
  \caption{Doppler images of $\sigma$\,Gem obtained for the seven data subsets. The x-axis is rotational phase, and the y-axis is spot latitude. The tick marks above each map denote the phases of the spectroscopic observations. The colour gives the temperature in Kelvin, as indicated by the colour scale on the bottom row. All the maps are plotted on the same temperature scale.}
  \label{DI}
\end{figure*}

\begin{figure*}
  \centering
  \includegraphics[width=18cm]{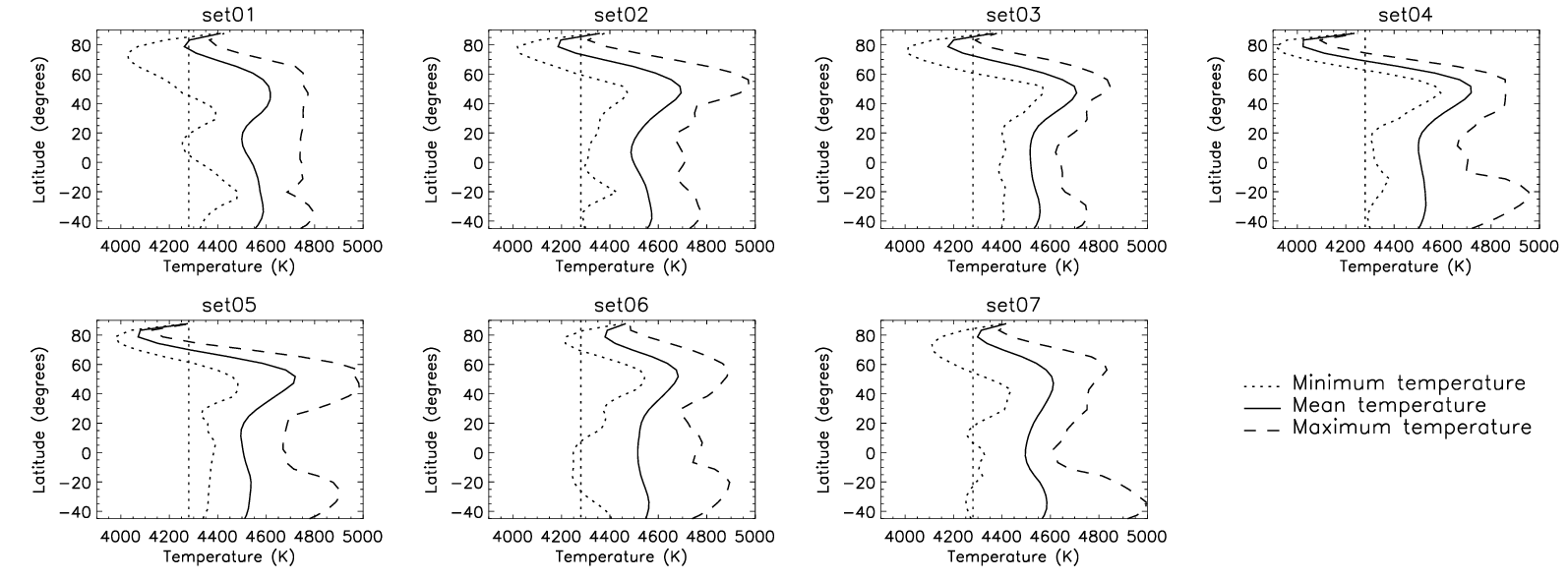}
  \caption{Minimum (dotted line), mean (solid line) and maximum (dashed line) temperature of the Doppler images plotted at each latitude. The vertical dotted line gives the temperature limit below which the area is considered to be a cool spot.}
  \label{DI_lat}
\end{figure*}

\begin{figure*}
  \centering
  \includegraphics[width=18cm]{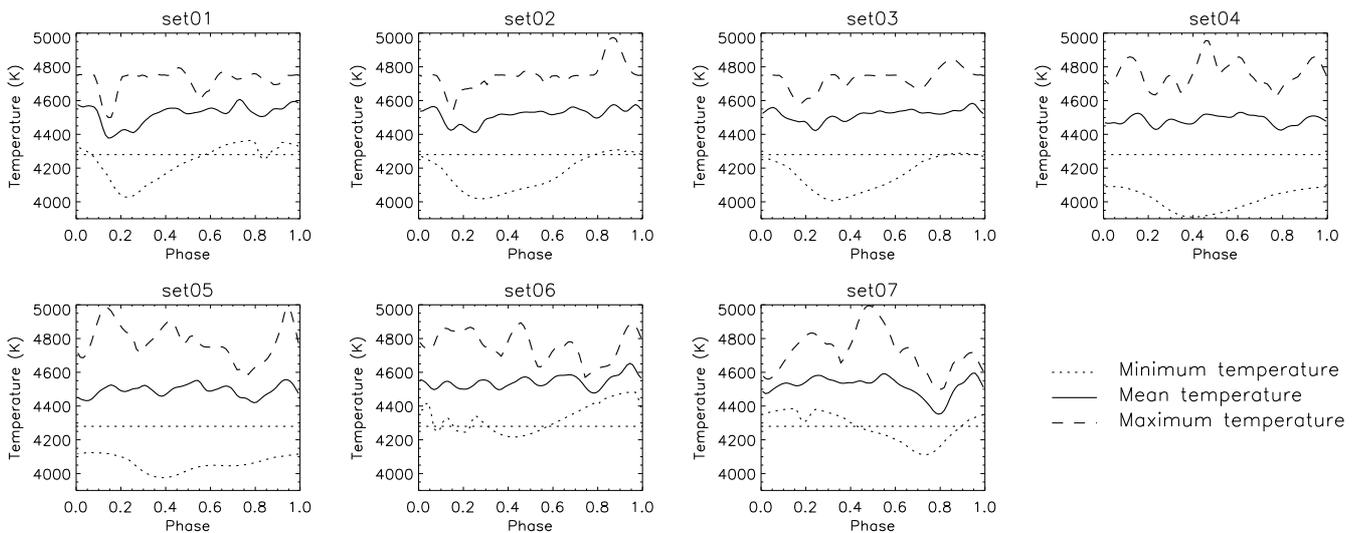}
  \caption{Minimum (dotted line), mean (solid line) and maximum (dashed line) temperature of the Doppler images plotted at each phase. The horizontal dotted line gives the temperature limit below which the area is considered to be a cool spot.}
  \label{DI_lon}
\end{figure*}

\section{Results}
\label{sect:results}

\subsection{Surface temperature maps}

Figure\,\ref{DI} shows the stellar surface temperature maps obtained with Doppler imaging techniques for all the subsets of \sG\ SONG data. All the maps show high-latitude spot features, often a polar spot with extensions to lower latitudes. High-latitude spots have also been recovered in \sG\ by other authors \citep{2015A&A...573A..98K,2017ApJ...849..120R}. The minimum spot temperature recovered from the SONG data is 3900\,K, again similar to previously published results \citep{2001A&A...373..199K,2015A&A...573A..98K,2017ApJ...849..120R}.

For a more detailed investigation of the spot locations the minimum, mean, and maximum temperature are plotted against the latitude (see Fig.\,\ref{DI_lat}) and the phase (see Fig.\,\ref{DI_lon}). We consider areas with the temperature of 250\,K less than that of the photosphere to be cool spots. On the Sun the penumbra has a temperature of 750--250\,K less than the photospheric one, depending on if it is dark or bright filament \citep[see, e.g.][]{2006A&A...453.1117B}. On the other hand the spot contrast on \sG\ is approximately half of the solar value of 1500\,K. Therefore, using spot limit of 250\,K less than that of the unspotted surface would correspond to the bright penumbral filaments on the Sun and also be close to half of the full spot contrast (equivalent of the solar penumbral dark filaments).

For the time period covered by the SONG data, November 2015 to April 2016, the cool spots on \sG\ concentrate at latitudes higher than 45$^{\circ}$. In the beginning of this time interval, namely in set01, the high-latitude spots concentrate around orbital phase 0.25, extending towards phase 0.5. During this interval, the spots reach down to latitude 45$^{\circ}$. Subsequently, the extensions towards later phase get more prominent, and, in set02 and set03, the spots at phase 0.25 start to lose their dominance. At the same time, the whole active region moves towards a higher latitude. In January 2016, set04 and set05, the spots cover more or less uniformly the whole polar region, reaching down only to latitude $\sim$60$^{\circ}$. In set06, the spots are seen at phases 0.3--0.6. However, the large phase gap between 0.5--0.9 makes the real extent of this high-latitude spot impossible to determine. Still, it is clear that spots do not cover phases 0.0--0.3 as they do in the earlier maps. In March 2016, set07, spots are seen mainly around orbital phase 0.75, and they are again extending towards lower latitudes, down to latitude $\sim$54$^{\circ}$. The minimum spot temperature is lowest during the interval when spots cover the whole polar region (set04 and set05) The temperature contrast between spotted and unspotted stellar surface is lowest in set06, but this could be an artefact due to sparse data coverage. 

In addition to the prominent spot activity seen at high latitudes, there are additional cool spots in the maps. They are typically smaller and have lower contrast than the high-latitude region. These spots appear to be smaller, short-lived active regions or individual spots and are seen in only one or two maps. The prominent cool spots that occur in some of the maps in the `southern' hemisphere, at the same phases as the most prominent cool spot regions in the `northern' hemisphere, are most likely artifacts (for example in set02 at phase 0.25 and latitude $-45^{\circ}$).

The surface maps also exhibit hot spots with temperatures up to 450\,K hotter than the surrounding photosphere. Hot spots can be artifacts due to poor phase coverage. For example, \cite{2014A&A...562A.139L} found that poor phase coverage increases the contrast of cool features and can create artificial hot spots at the same phase. In the \sG\ maps presented here, the phase coverage is mostly very good, but there are a few instances where a hot spot is paired with a cool spot when the phase coverage is not optimal. For example, the most prominent hot spot in set02 occurs around phase 0.85, corresponding to the largest phase gap in our observations. It is also paired with lower-latitude cool spots. However, this cool feature is also seen in set01 and weakly in set03. Therefore, at least the cool spot in set02 is probably real, but its shape, size, and temperature are poorly constrained. The large hot spot around phase 0.5 and latitude $-40^{\circ}$ is seen in every map but with changing temperature contrast and morphology. Our phase coverage is very good around this phase, so we think this hot spot is real. 

\begin{figure*}
  \centering
  \includegraphics[width=18cm]{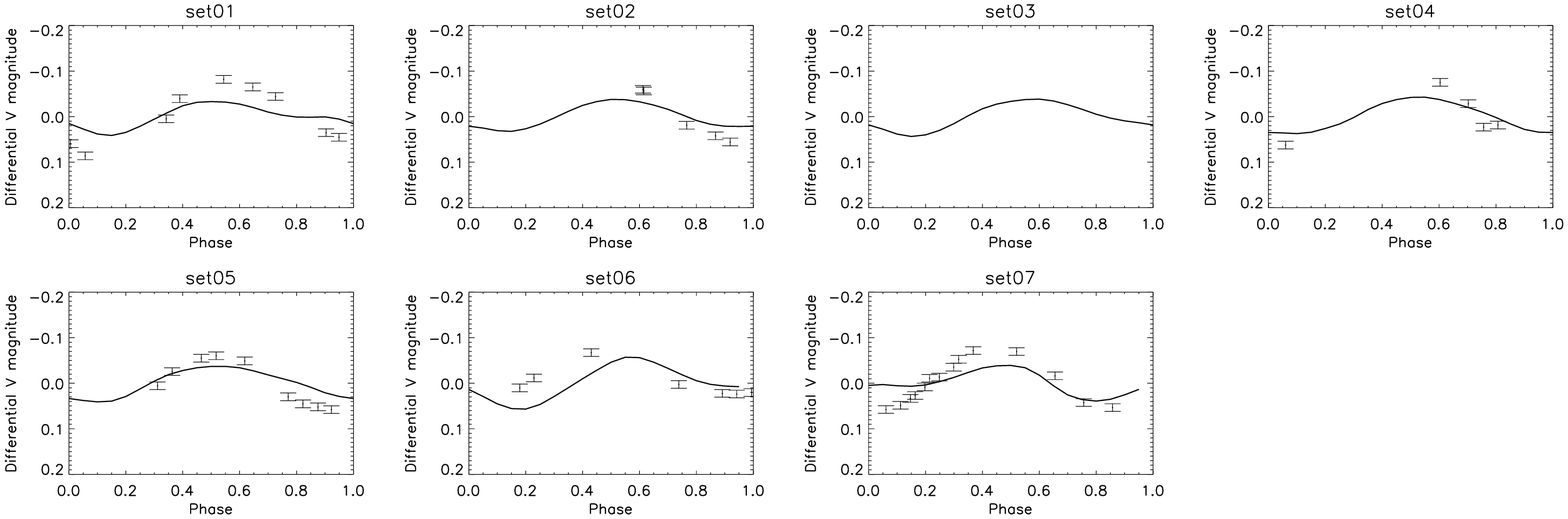}
  \caption{Normalised {\it V}-band magnitudes calculated from the temperature maps (line) compared to contemporaneous observations. The error bars in the observed photometry reflect the typical error of the observations, not the error bars of individual data points.}
  \label{photom_sets}
\end{figure*}

To check the reliability of our Doppler images, we compare the brightness of \sG\ computed from the Doppler images with the $V$-band photometric observations. Figure\,\ref{photom_sets} shows the computed light curves (solid lines) plotted against the normalised contemporaneous differential {\it V} magnitudes (symbols with error bars). In general, the shape of the light curves calculated from the Doppler images agree well with the photometric observations. However, it appears in general that the light curves computed from the Doppler images underestimate the amplitude of the observed light curves. A similar result was obtained by \cite{2017ApJ...849..120R}, who found the amplitude computed from the Doppler images was approximately 70\% of the observed amplitude. We find a similar difference, but exact comparisons are difficult at times due to gaps in the photometric observations. 

\subsection{Chromospheric activity}

In the chromosphere, solar magnetic activity exhibits itself as bright regions of excess emission (plages) and also as prominences that are formed by material trapped in magnetic loops that extend into the corona. Plages are often correlated with large sunspots on the solar photosphere, but they can also occur without spots. For a review of the evolution of active regions in different layers of the solar atmosphere, see \cite{2015LRSP...12....1V}. In stars other than the Sun, Ca H\&K and H$\alpha$ spectral lines are often used as a proxy for chromospheric activity. Unfortunately, the SONG spectral range does not extend to the Ca H\&K spectral lines, but H$\alpha$ is covered.

In \sG\ the H$\alpha$ line is in absorption, and it shows clear variability. This variability is caused by the changing contribution from the plages in the chromosphere. The increasing emission from the plages partially fills in the spectral line, making it weaker. The H$\alpha$ line profiles can also show evidence for cool prominences extending to the stellar corona. They would be seen as enhanced emission outside the stellar disc and as absorption features when seen against the disc.  

\begin{figure*}
  \centering
  \includegraphics[width=18cm]{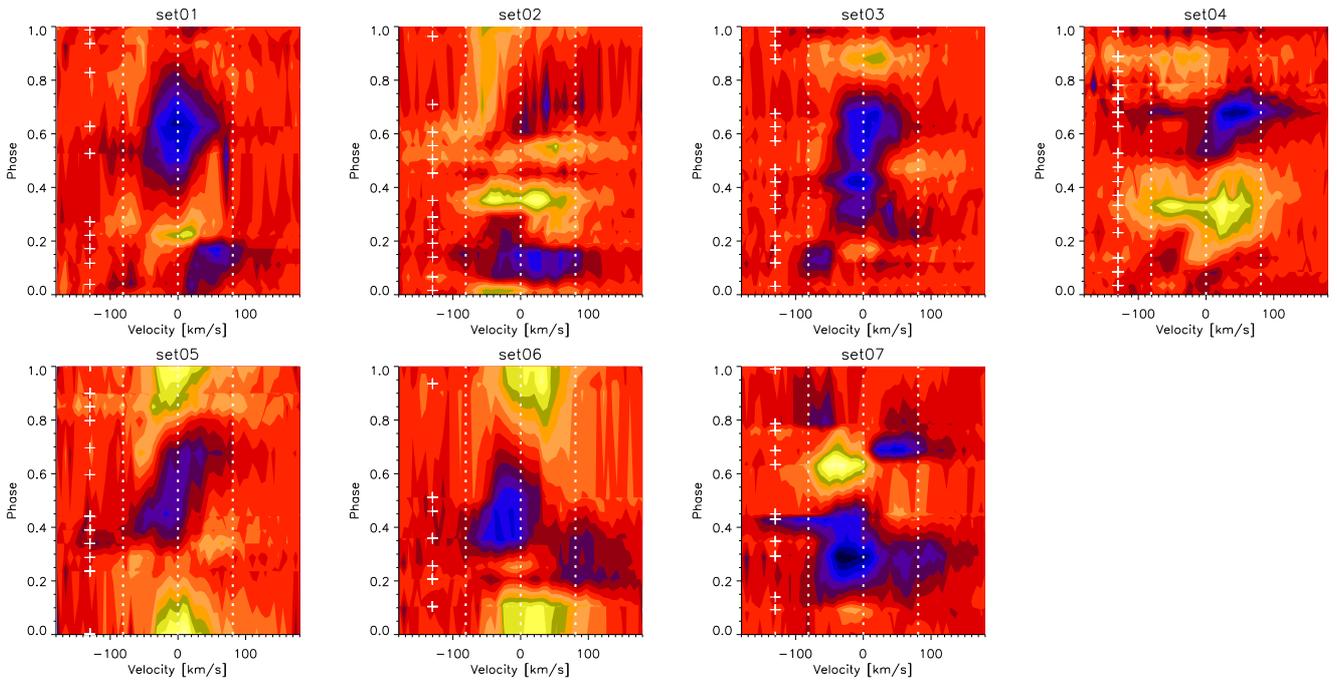}
  \caption{Dynamic H$\alpha$ residual spectra of $\sigma$~Gem with the average line profile subtracted. The yellow denotes areas of increased emission and the blue areas of enhanced absorption in comparison to the average level. Crosses on the left show the phases of individual spectra. The dotted lines mark the blue shifted line wing, line core, and the redshifted line wing of the H$\alpha$ line profile.}
  \label{Ha_dynamic}
\end{figure*}

To study time variability in the H$\alpha$ line in detail, dynamic spectra were constructed from the line profiles. To enhance the differences between them, an average line profile constructed from all 256 spectra was subtracted from the individual H$\alpha$ line profiles. The dynamic spectra are shown in Fig.\,\ref{Ha_dynamic}. The bright yellow areas denote enhanced emission and dark blue areas increased absorption in comparison to the average level. The orbital phases with observations are indicated with white crosses on the left side of each panel. Wherever data points are missing, we interpolated between the orbital phases with data. The dynamic spectra show both regions of enhanced emission and increased absorption.

The three vertical dotted lines in Fig.\,\ref{Ha_dynamic} give the location of the blue shifted line wing, line core, and the redshifted line wing. It is clear that the enhanced emission and absorption features are seen at velocities that coincide with the stellar disc. We identify the following main absorption features: phases $0.5-0.8$ in set01, around phase 0.1 in set02, at phases $0.3-0.7$ in set03, at phases $0.5-0.7$ in set04, at phases $0.4-0.7$ in set05, at phases $0.3-0.6$ in set06, and phases $0.2-0.4$ in set07. The enhanced emission occurs at phase 0.2 in set01, phase 0.4 in set02, around phase 0.3 in set04, at phases $0.8-0.1$ in set05, at phases $0.9-0.1$ in set06, and around phase 0.6 in set07. We also note that the set06 has a phase gap extending phases $0.5-0.9$. On the whole, the chromospheric features seem to be quite short lived and are rarely seen to move in the line profile due to the stellar rotation. This is true even in the cases where the phase coverage is very good.

\begin{figure*}
  \centering
  \includegraphics[width=18cm]{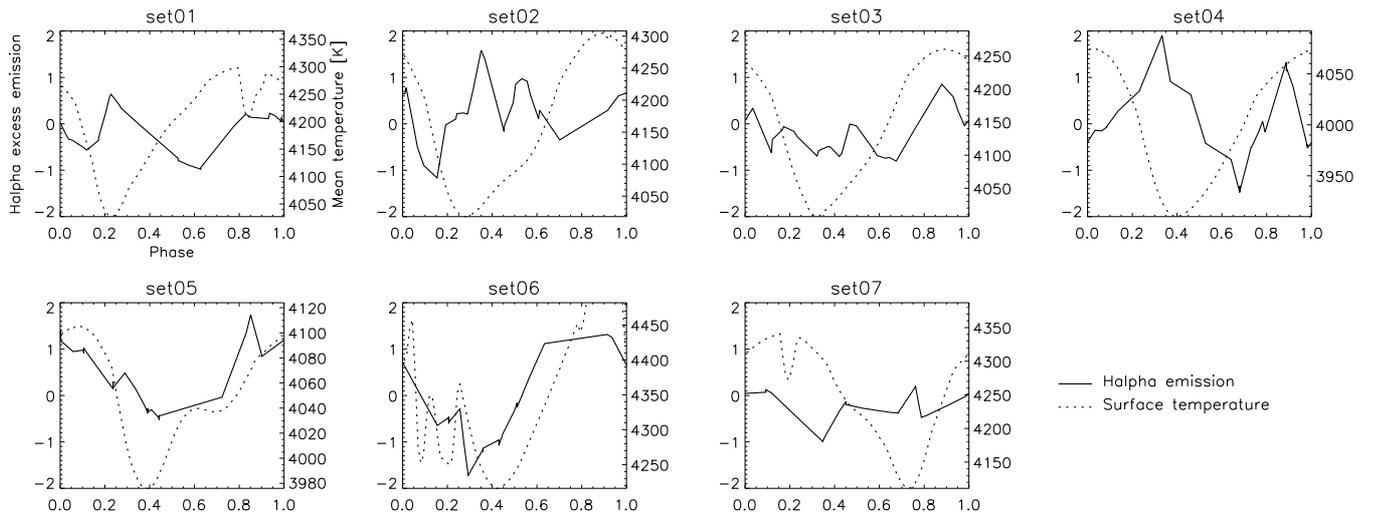}
  \caption{Comparison of the photospheric and chromospheric activity. Dotted lines show the minimum temperature and the solid lines the residual H$\alpha$ emission plotted against the phase.}
  \label{Ha_spots}
\end{figure*}

On the Sun, the chromospheric emission originating from plage regions is often correlated with nearby cool sunspots on the photosphere \citep[e.g.][]{2015LRSP...12....1V}. Similar correlations have been detected on other stars as well \citep[e.g.][]{2005AJ....130.1231P,2008MNRAS.385..708S}, but this correlation is not always observable.  When it is detected, it is often weak. To search for possible correlations between photospheric and chromospheric activity on \sG , the minimum temperature in Doppler images is compared to the summed H$\alpha$ excess emission at different phases. 

In general, as can be seen from Fig.\,\ref{Ha_spots}, no clear correlation between the two measures is seen. There are a few times when the minimum surface temperature (dotted line) corresponds to a high H$\alpha$ excess emission (solid line), such as set01 around phase 0.2, set02 around phases $0.2-0.4$, set04 around phases $0.2-0.4$, and set05 at phase 0.8. This implies that cool photospheric spots on \sG\ are sometimes associated with plages in the chromosphere. The correlation is not strong and is not seen at all times. However, our current observations are only snapshots in time, and higher time resolution observations would help in studying this behaviour in more detail. 

While the continuum normalisation is done as carefully as possible, the region closest to the H$\alpha$ profile is still allowed to vary. The continuum normalisation points near the H$\alpha$ line core are at 6559.8\AA\,and 6569.0\AA. Outside this region the continuum normalisation is very good, but variations are seen inside the region. Some of these variations are undoubtedly real stellar variations, but some may be instrumental effects. However, this does not change the orbital phases at which the main regions of increased absorption and enhanced emission occur.

\section{Discussion}
\label{sect:disc}

\subsection{Spot evolution}
\label{sect:evolution}

The Doppler images of \sG\ obtained from continuous SONG data offer a unique opportunity to study detailed spot evolution during the 150-night-long observing period. Our Doppler map obtained from set01 shows the majority of cool spot activity to be concentrated at phase 0.25, whereas the last map, set07, shows the spots concentrated around phase 0.75. This could indicate a so-called flip-flop event, where the largest spot region abruptly changes longitude by 180$^{\circ}$.

The flip-flop phenomenon was first discovered in photometry of the single active G giant FK\,Com by \cite{1993A&A...278..449J}. Since then, it has been further studied on FK\,Com and also reported on many other active stars \citep[see, e.g.][]{1998A&A...336L..25B,2002A&A...390..179K,2005A&A...432..657J}, including the Sun \citep{2003A&A...405.1121B}. The flip-flop phenomenon has also been reported on \sG\ by \cite{1998A&A...336L..25B}. They found that spots on \sG\ concentrate at two active longitudes that are 180$^{\circ}$ apart and that the regions of greatest activity switch between the two active longitudes every few years. However, more recent studies have shown that the picture is more complicated. The changes in active longitudes are often not regular and can occur in `phase jumps' instead of the full 180$^{\circ}$ flip-flops \citep[see, e.g.][]{2006A&A...452..303O,2013A&A...553A..40H,2014A&A...562A.107K}. 

The change in spot configuration seen in our SONG temperature maps of set01 and set07 would appear to be a classical flip-flop event, if the maps between set01 and set07 were missing. Most flip-flops reported in the literature have been detected through broadband photometric observations, therefore we investigate how the current event looks from our photometric observations alone. Our photometry is quite sparse, but we can still study the observations we have and compare them to the calculated light curves based on our temperature maps. As can be seen in Fig.\,\ref{photom_sets}, the light curve minimum remains around phase 0.15--0.20 throughout the interval mapped by set01--set05. Only small variations in the location of the minimum and the shape of the light curves are seen. The light curve from set07 shows that the minimum has shifted to phase of 0.8 (calculated from the map) or 0.9 (photometric observations). The light curve calculated from set06 displays features intermediate between set01--05 and set07. The light curve of set06 has a wider minimum, but it remains centred around phase 0.2, and a minimum around phase 0.8 is becoming more prominent. The observed light curve of set06 is not well fit around phases 0.2--0.4 by the light curve calculated from the Doppler image.

\begin{figure}
  \centering
  \includegraphics[width=9cm]{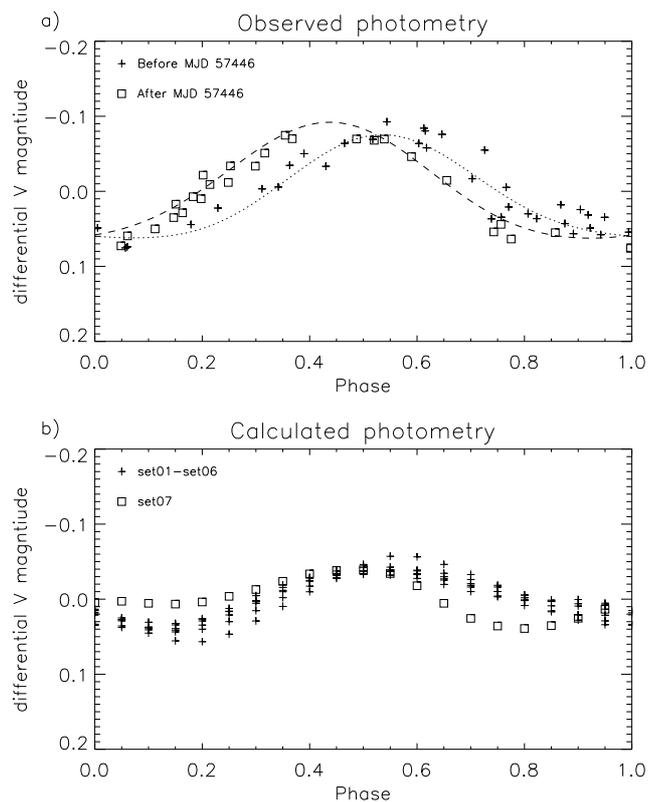}
  \caption{Comparison of the observed brightness (a) with the brightness computed from the Doppler images (b). In both cases the observations before the phase jump, which occurred around MJD=57446, are denoted with plus-signs. The observations after the phase jump are marked with squares. In a) the dotted line is the harmonic fit to the observations before the phase jump and the dashed line the harmonic fit to the observations after the phase jump.}
  \label{jump}
\end{figure}

The spot evolution described above would not be considered a flip-flop event, but possibly a phase jump \citep{2006A&A...452..303O,2013A&A...553A..40H}. Studying the observed photometry in detail reveals that the phase jump occurs around  MJD=57446. This corresponds with a gap in our spectroscopic observations, and the photometric observations are too sparse to pinpoint the time of the event. Fig.\,\ref{jump} shows both the photometric observations and the brightness calculated from the temperature maps, but now with different symbols for the phases before (plus signs) and after (empty squares) the jump. Both the observed and the calculated brightness show the maximum moving 0.10--0.15 in phase towards earlier phases and the minimum around phase 0.7--0.8 becoming more pronounced. To quantify this event we did a harmonic fit, using the base and one harmonic, to the whole photometric data set simultaneously and to the two sets obtained before and after the proposed phase jump separately (fits to the two subsets are plotted in Fig.\,\ref{jump}a). The fits show a maximum at the phase 0.54 before the event and at the phase 0.44 after the event. Similarly, the minimum is at the phase 0.07 before the event and at the phase 0.92 after the event. Additionally, when fitting the whole data set simultaneously the residuals have a standard deviation that is at least 50\% larger than when using the two separate fits.

One has to keep in mind that the phases are calculated from the orbital period, not the rotation period determined from the spots themselves. The spot rotation period for this time period is somewhat shorter than the orbital period, making the spots move slowly towards later phases with time. However, this would not explain a sudden jump such as the one we see here. With the data presented here, we cannot confirm whether the jump in phases results in a long-term change in activity location. However, photometry obtained after the SONG data, during the time period $MJD=57480 - 57513$ (2 April -- 3 May 2016), shows the light curve shape to be nearly identical to the time period of set07. These data points are also plotted in Fig.\,\ref{jump}.

\begin{figure}
  \centering
  \includegraphics[width=9cm]{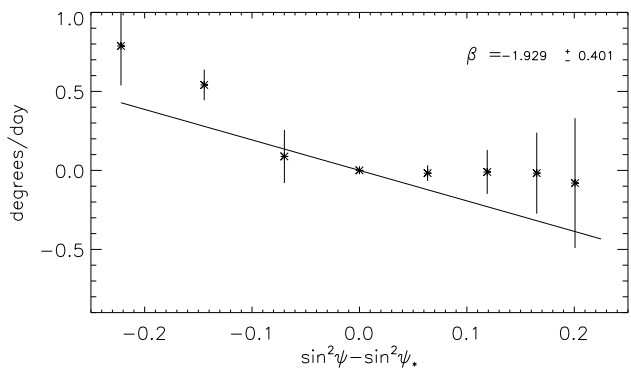}
  \caption{Phase shifts between the five temperature maps obtained from the SONG data for set01--set05. The constant movement of the spots has been removed from the measurements, by removing the shift at the lowest latitude ($\psi_{*}=60.25^{\circ}$) that always has spots. The plot gives the average phase shifts in degrees per day plotted against $\sin^{2}(\psi)-\sin^{2}(\psi_{*})$. The error bars indicate the standard deviation. The weighted fit of the function $f(\psi)=\beta(\sin^{ 2}(\psi)-\sin^{2}(\psi_{*}))$ to the measurements is shown as a straight line.
  }
  \label{DR}
\end{figure}

\subsection{Surface differential rotation}

The seven temperature maps obtained from the SONG data covering eight consecutive stellar rotations present a unique opportunity to study surface differential rotation on \sG. These maps can be used for a cross-correlation investigation, where the individual latitude strips from two consecutive maps are cross-correlated and the measured spot shifts are analysed to obtain the surface differential rotation.

The phase jump occurring between set06 and set07 and the associated fast spot evolution make it impossible to study differential rotation in this time interval. Additionally, set06 has a large data gap at phases 0.5--0.9, so it is not possible to determine the real extent of the high-latitude spot. Therefore, we use only the first five maps (set01--set05) for the further analysis of surface differential rotation.

The relative surface differential rotation coefficient is defined as:
\begin{equation}
  \alpha=\frac{\Omega_{0}-\Omega_{P}}{\Omega_{0}},
  \label{eq:alpha}
\end{equation}
 where $\Omega_{0}$ and $\Omega_{P}$ are the equatorial and polar angular velocities, respectively. Unfortunately, there are no spots near the equator. Therefore, a comparison latitude $\psi_{*}$ other than the equator has to be used. With a $\sin^{2}\psi$ law of the surface differential rotation, the angular velocity at any latitude can be expressed as:
\begin{equation}
  \Omega = \Omega_{0} + \beta \sin^{2}\psi.
  \label{eq:lat}
\end{equation}
The polar angular velocity can be obtained from Eq.\,\ref{eq:alpha} and $\alpha$ can then be represented as:
\begin{equation}
  \alpha=\frac{-\beta}{\Omega_{0}}.
  \end{equation}
We can determine $\beta$ by determining how the spots move at different latitudes. For this we correlated all the latitude strips that had spots, excluding the pole. This gives usable measurements from latitude +45$^{\circ}$ to latitude +85.5$^{\circ}$.

As mentioned before, the phases are calculated from the orbital period, not the rotation period measured from the spots. The spot rotation period is somewhat shorter than the orbital period, and it also changes with time depending on the latitude at which the dominant spots are located. Therefore, before analysing the cross-correlation results, the constant shift caused by the difference in the orbital and rotation periods is removed from the measurements. This is done using the shifts at the lowest latitude that always has spots, meaning 60.75$^{\circ}$ (centre of the 4.5$^{\circ}$ wide latitude strip). The shift measurements per latitude strip obtained from different maps are averaged, and the standard deviation of the measurements is used as the error for that latitude strip. The results obtained are plotted in Fig.\,\ref{DR}, and the value of $\beta$ is determined by using a first order polynomial fit with the inverse standard deviation of the average in each point as the weight. The fit resulted in $\beta=-1.9\pm0.4^{\circ}/$day. The errors for high latitudes are large because the spot distribution is almost constant at these latitudes. The highest latitude used in the analysis, 85.5$^{\circ}$, is missing from Fig.\,\ref{DR} because its error bar would extend over the whole range of the plot and even beyond. If one discards the measurements from the highest latitudes from the fit and only uses latitudes from $+45^{\circ}$  to $+67.5^{\circ}$, the resulting $\beta$ is the same within the errors, $\beta=-2.4\pm0.5^{\circ}/$day. This implies solar-like surface differential rotation somewhat weaker than measured for the Sun \cite[$\beta=-2.9$ by][]{1986A&A...155...87B}.

For the calculation of $\alpha$, the angular velocity at the equator is still needed. That can be calculated from Eq.\,\ref{eq:lat} after the velocity at the comparison latitude is known. Determining the rotation period at the comparison latitude that always has spots, $\psi_{*}=60.75^{\circ}$, can be done using photometry. The rotation period determined from the photometric observations best reflects the rotation period at the most visible latitude with spots, in other words the lowest latitude with the spots. Naturally the value determined this way is also affected by the spot motion on the other latitudes, but it is the most reasonable estimate. The $V$-band photometry gives a rotation period $P_{rot}=19.25\pm 0.14$\,days. This translates to angular velocity $\Omega_{*}=18.70\pm0.14^{\circ}$/day and equatorial rotation period of $17.9\pm0.6$ days \citep[similar to the shortest periods determined by][]{2014A&A...562A.107K}. Together with $\beta$ this yields $\alpha=0.10\pm0.02$. This implies relative strong solar-like differential rotation.  

This result is well compatible with $\alpha=0.103$ obtained by \cite{2014A&A...562A.107K}. They used long-term photometric observations to investigate the variations in the stellar rotation period. They caution that their result could be caused by insufficient sampling in their data. Other studies have found much weaker solar-like differential rotation on \sG\ \citep{1995ApJS...97..513H} or anti-solar differential rotation \citep{2007A&A...474..165K, 2015A&A...573A..98K}. On other RS\,CVn-type binaries, varying results have been obtained concerning surface differential rotation. In general, Doppler imaging studies have yielded weaker solar-like surface differential rotation or anti-solar rotation on cool giant stars \citep[for a recent review see][]{2017AN....338..903K}. The strength of the surface differential rotation measured here for \sG\ is about twice that of other giants with similar rotation periods. On the other hand, the results based on Doppler imaging also strongly depend on the time span between maps and possible other spot evolution present at the time. Here continuous observations are used making the determination more reliable. This is also the case in some, but not all, of the results used by \cite{2017AN....338..903K}.

We want to caution that these results are based on spots on a very limited latitude range. This can easily cause errors in the exact determination of the strength of the surface differential rotation. One should also note that it is not certain that large starspots, such as the ones on \sG, are good tracers of surface differential rotation. Based on dynamo simulations, \cite{2011A&A...532A.106K} showed that stellar spots caused by the large-scale dynamo fields are not necessarily tracing the stellar differential rotation, whereas spots formed from small-scale fields trace the surface flow patterns well. One can wonder whether the large spots on \sG\ can be caused by small scale fields. According to \cite{2011A&A...532A.106K}, this should lead to more solid-body-like rotation, not to large surface differential rotation. Additionally, this investigation relies on the large, high-latitude spots to exist long enough to be shaped by the differential rotation and also the same spots being present in consecutive maps. These are reasonable assumption since the spots on evolved stars such as \sG\ are thought to be longer-lived than spots in young active stars \citep[see, e.g.][]{2002AN....323..349H,2007A&A...464.1049I}.

\section{Conclusions}
\label{sect:conc}

In this paper we have investigated continuous spectroscopic and photometric observations of \sG\ obtained in 2015--2016. The spectroscopic observations cover eight continuous stellar rotations and allow for detailed study of photospheric and chromospheric activity during this interval. Our investigation includes the study of spot evolution and surface differential rotation, as well as the correlation of photospheric and chromospheric activity. We draw the following main conclusions:

\begin{enumerate}
\item During the 2015--16 observing season, \sG\ has shown primarily high-latitude and polar spots at latitudes higher than 45$^{\circ}$ and usually higher than 60$^{\circ}$.
\item The spots seen in the temperature maps concentrated around phase 0.25 at the beginning of the observing period and moved to phase 0.75 by the end of our observations. In between these two extremes, the spots were almost polar, with a few extensions to slightly lower latitudes.
\item The photometric observations and the brightness calculated from the temperature maps confirm the change in the spot phases. However, the changes are much smaller than when comparing just the larger spots seen in the temperature maps. The photometry confirms a phase jump occurring in late February 2016.
\item The consecutive temperature maps obtained before the phase jump were cross-correlated to study surface differential rotation. We find a solar-like differential rotation law with the equator rotating faster than the polar regions. The relative differential rotation coefficient is $\alpha=0.10\pm0.02$. This result implies much stronger surface differential rotation than in most of the earlier investigations, some of which have also yielded results with anti-solar differential rotation. 
\item Chromospheric activity was investigated from $H\alpha$ line profiles. No clear flares were seen in the data, but the line profile is highly variable, mainly due to plages.
\item The comparison between photospheric and chromospheric activity shows that the plages are at times associated with photospheric spot activity. However, this correlations is weak and not always seen.
\end{enumerate}

For further studying the spot evolution and pinpointing the surface differential rotation on \sG\ more continuous spectroscopic observations are needed. This would be especially beneficial during time periods when lower latitude spots are also present enabling more accurate determination of the strength of the surface differential rotation.

\begin{acknowledgements}
We would like to thank the anonymous referee for their comments that helped to improve the paper. Based on observations made with the Hertzsprung SONG telescope operated on the Spanish Observatorio del Teide on the island of Tenerife by the Aarhus and Copenhagen Universities and by the Instituto de Astrofisica de Canarias. RMR acknowledges support from the Yale Center for Astronomy \& Astrophysics (YCAA) Prize Postdoctoral Fellowship. SG would like to acknowledge support from the National Natural Science Foundation of China (grant Nos. 10373023, 10773027, 11333006, and U1531121). GWH acknowledges long-term support from NASA, NSF, Tennessee State University, and the State of Tennessee Centers of Excellence program. VA was supported by a research grant (00028173) from VILLUM FONDEN. Funding for the Stellar Astrophysics Center is provided by The Danish National Research Foundation (Grant agreement no. DNRF106). This work has made use of the SIMBAD database at CDS, Strasbourg, France, and NASA's Astrophysics Data System (ADS) services.
\end{acknowledgements}

%
%

\bibliographystyle{aa} 
\bibliography{sigmaGem_SONG_astroph} 

\begin{appendix} 

  \section{Separating set07 into two individual maps}
\label{set07}  

The data used for set07 covers 1.6 stellar rotations. To check that no significant changes in the spot configuration occurred during this time period, we have divided the data into two subsets and imaged them separately. The data are divided at MJD=57464, with the observations obtained that night being used in both subsets.
  
\begin{figure}
  \centering
  \includegraphics[width=9cm]{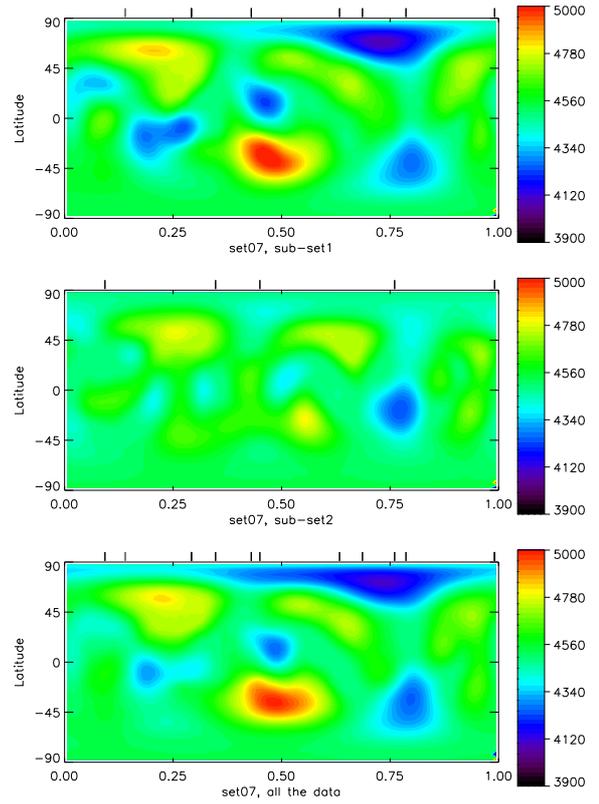}
  \caption{Doppler images of $\sigma$~Gem obtained from set07 divided into two subsets at MJD=57464 and using all the data together (set07 in our analysis). The data for subset1 were obtained at MJD=57448--57464 and for the subset2 at MJD=57464--57479. The x-axis is rotational phase, and the y-axis is spot latitude. The tick marks above each map denote the phases of the spectroscopic observations. The colour gives the temperature in Kelvin, as indicated by the colour scale next to the map. All the maps are plotted on the same temperature scale.}
  \label{sets7_8}
  \end{figure}

 In Fig.\,\ref{sets7_8} the maps obtained from the two subsets and the map from the combined set07 are shown. As can be seen, the maps from both of the subsets have the main spot at the same phase as in combined set07. The spot longitude is well recovered in both cases -- also for the subset2 due to having data from the same phase as the spot. Still, the latitude information of the spot is lost in the map obtained from subset2. The closest phases with spectra on both sides of the spot are some 0.2 phase units away, giving extremely limited information on the spot latitude. Therefore, the spot latitude is not reliably recovered. Still, the maps obtained from the two subsets show similar spot locations as the combined map, and therefore we are confident that we can use all these data in our analysis.

  In Fig.\,\ref{sets7_8} the temperatures are only plotted up to 5000\,K, meaning that the ten surface points in the map obtained from the subset1 that originally had temperatures between 5000\,K and 5058\,K are now set equal to 5000\,K. This is done in order to have the same temperature scale for all maps presented in this paper. All the spots found in the maps we use for our analyses have maximum temperature values lower than 5000\,K. 

  \section{Observing logs}

The full observing logs for the photometric observations are given in Table\,\ref{obs_log_photom} and for the spectroscopic observations in Table\,\ref{obs_log_spec}.
    

  \longtab[1]{
\begin{longtable}{cccccc}
\caption{Photometric observing log.}\\
\hline
\hline
Date & MJD & Orbital & Filter & differential & ellipticity \\
     &  +57000 & Phase &  & magnitude  & removed \\
\hline
\endfirsthead
\caption{Continued.} \\
\hline
Date & MJD & Orbital & Filter & differential & ellipticity \\
     & +57000 & Phase &  & magnitude & removed  \\
\hline
\endhead
\hline
27-10-2015 & 322.4108 & 0.726 & V & -1.086 & -1.097 \\
27-10-2015 & 322.4108 & 0.726 & B & -0.977 & -0.988 \\
08-11-2015 & 334.4823 & 0.342 & V & -1.041 & -1.047 \\
08-11-2015 & 334.4823 & 0.342 & B & -0.927 & -0.934 \\
09-11-2015 & 335.4168 & 0.389 & V & -1.095 & -1.092 \\
12-11-2015 & 338.4438 & 0.544 & V & -1.145 & -1.134 \\
14-11-2015 & 340.4494 & 0.646 & V & -1.114 & -1.117 \\
14-11-2015 & 340.4494 & 0.646 & B & -1.012 & -1.016 \\
19-11-2015 & 345.5069 & 0.904 & V & -1.021 & -1.017 \\
19-11-2015 & 345.5069 & 0.904 & B & -0.901 & -0.897 \\
20-11-2015 & 346.3907 & 0.949 & V & -1.017 & -1.007 \\
20-11-2015 & 346.3907 & 0.949 & B & -0.896 & -0.885 \\
21-11-2015 & 347.4869 & 0.005 & V & -1.007 & -0.993 \\
21-11-2015 & 347.4869 & 0.005 & B & -0.888 & -0.872 \\
22-11-2015 & 348.5008 & 0.057 & V & -0.978 & -0.966 \\
22-11-2015 & 348.5008 & 0.057 & B & -0.852 & -0.839 \\
03-12-2015 & 359.3927 & 0.612 & V & -1.127 & -1.126 \\
03-12-2015 & 359.3927 & 0.612 & B & -1.022 & -1.021 \\
03-12-2015 & 359.4476 & 0.615 & V & -1.123 & -1.122 \\
03-12-2015 & 359.4476 & 0.615 & B & -1.022 & -1.021 \\
06-12-2015 & 362.4007 & 0.766 & V & -1.036 & -1.047 \\
06-12-2015 & 362.4007 & 0.766 & B & -0.922 & -0.934 \\
08-12-2015 & 364.3930 & 0.868 & V & -1.022 & -1.024 \\
08-12-2015 & 364.3930 & 0.868 & B & -0.897 & -0.899 \\
09-12-2015 & 365.3952 & 0.919 & V & -1.016 & -1.010 \\
09-12-2015 & 365.3952 & 0.919 & B & -0.895 & -0.888 \\
11-01-2016 & 398.4166 & 0.603 & V & -1.108 & -1.105 \\
11-01-2016 & 398.4166 & 0.603 & B & -1.009 & -1.006 \\
12-01-2016 & 399.4458 & 0.656 & B & -0.966 & -0.971 \\
13-01-2016 & 400.3658 & 0.703 & V & -1.049 & -1.058 \\
13-01-2016 & 400.3658 & 0.703 & B & -0.933 & -0.943 \\
14-01-2016 & 401.4236 & 0.757 & V & -0.996 & -1.007 \\
14-01-2016 & 401.4236 & 0.757 & B & -0.898 & -0.910 \\
15-01-2016 & 402.4063 & 0.807 & V & -1.003 & -1.012 \\
15-01-2016 & 402.4063 & 0.807 & B & -0.877 & -0.886 \\
18-01-2016 & 405.3108 & 0.955 & B & -0.879 & -0.867 \\
20-01-2016 & 407.3774 & 0.060 & V & -0.979 & -0.968 \\
20-01-2016 & 407.3774 & 0.060 & B & -0.862 & -0.849 \\
25-01-2016 & 412.2952 & 0.311 & V & -1.033 & -1.045 \\
25-01-2016 & 412.2952 & 0.311 & B & -0.929 & -0.943 \\
26-01-2016 & 413.3078 & 0.363 & V & -1.074 & -1.076 \\
26-01-2016 & 413.3078 & 0.363 & B & -0.959 & -0.961 \\
28-01-2016 & 415.3109 & 0.465 & V & -1.119 & -1.105 \\
28-01-2016 & 415.3109 & 0.465 & B & -1.015 & -1.000 \\
29-01-2016 & 416.3410 & 0.518 & V & -1.124 & -1.111 \\
29-01-2016 & 416.3410 & 0.518 & B & -1.023 & -1.008 \\
31-01-2016 & 418.3032 & 0.618 & V & -1.100 & -1.099 \\
31-01-2016 & 418.3032 & 0.618 & B & -0.985 & -0.984 \\
03-02-2016 & 421.3020 & 0.771 & V & -1.010 & -1.021 \\
03-02-2016 & 421.3020 & 0.771 & B & -0.881 & -0.892 \\
04-02-2016 & 422.3210 & 0.823 & V & -0.998 & -1.005 \\
04-02-2016 & 422.3210 & 0.823 & B & -0.879 & -0.887 \\
05-02-2016 & 423.3519 & 0.875 & V & -0.998 & -0.999 \\
05-02-2016 & 423.3519 & 0.875 & B & -0.882 & -0.883 \\
06-02-2016 & 424.2782 & 0.922 & V & -0.999 & -0.993 \\
06-02-2016 & 424.2782 & 0.922 & B & -0.884 & -0.877 \\
10-02-2016 & 428.3071 & 0.128 & B & -0.849 & -0.849 \\
11-02-2016 & 429.3079 & 0.179 & V & -0.987 & -0.997 \\
11-02-2016 & 429.3079 & 0.179 & B & -0.865 & -0.877 \\
12-02-2016 & 430.2870 & 0.229 & V & -1.003 & -1.019 \\
12-02-2016 & 430.2870 & 0.229 & B & -0.888 & -0.907 \\
16-02-2016 & 434.2322 & 0.430 & V & -1.085 & -1.075 \\
16-02-2016 & 434.2322 & 0.430 & B & -1.001 & -0.989 \\
22-02-2016 & 440.2740 & 0.738 & V & -0.994 & -1.005 \\
22-02-2016 & 440.2740 & 0.738 & B & -0.881 & -0.893 \\
23-02-2016 & 441.2611 & 0.789 & B & -0.852 & -0.863 \\
25-02-2016 & 443.2663 & 0.891 & V & -0.987 & -0.985 \\
25-02-2016 & 443.2663 & 0.891 & B & -0.862 & -0.860 \\
26-02-2016 & 444.2636 & 0.942 & V & -0.993 & -0.984 \\
26-02-2016 & 444.2636 & 0.942 & B & -0.874 & -0.864 \\
27-02-2016 & 445.2830 & 0.994 & V & -1.001 & -0.987 \\
27-02-2016 & 445.2830 & 0.994 & B & -0.879 & -0.864 \\
01-03-2016 & 448.2789 & 0.147 & V & -1.004 & -1.008 \\
01-03-2016 & 448.2789 & 0.147 & B & -0.875 & -0.879 \\
02-03-2016 & 449.2797 & 0.198 & V & -1.020 & -1.033 \\
02-03-2016 & 449.2797 & 0.198 & B & -0.891 & -0.906 \\
03-03-2016 & 450.2726 & 0.249 & V & -1.038 & -1.055 \\
03-03-2016 & 450.2726 & 0.249 & B & -0.919 & -0.939 \\
04-03-2016 & 451.2604 & 0.299 & V & -1.063 & -1.077 \\
04-03-2016 & 451.2604 & 0.299 & B & -0.944 & -0.960 \\
11-03-2016 & 458.2460 & 0.655 & V & -1.053 & -1.058 \\
11-03-2016 & 458.2460 & 0.655 & B & -0.931 & -0.936 \\
13-03-2016 & 460.2251 & 0.756 & V & -0.988 & -0.999 \\
13-03-2016 & 460.2251 & 0.756 & B & -0.867 & -0.879 \\
15-03-2016 & 462.2084 & 0.857 & V & -0.985 & -0.988 \\
15-03-2016 & 462.2084 & 0.857 & B & -0.853 & -0.856 \\
19-03-2016 & 466.1963 & 0.061 & V & -0.995 & -0.984 \\
19-03-2016 & 466.1963 & 0.061 & B & -0.872 & -0.859 \\
20-03-2016 & 467.2030 & 0.112 & V & -0.996 & -0.993 \\
20-03-2016 & 467.2030 & 0.112 & B & -0.867 & -0.863 \\
21-03-2016 & 468.2068 & 0.163 & V & -1.007 & -1.014 \\
21-03-2016 & 468.2068 & 0.163 & B & -0.880 & -0.888 \\
22-03-2016 & 469.2037 & 0.214 & V & -1.037 & -1.052 \\
22-03-2016 & 469.2037 & 0.214 & B & -0.912 & -0.929 \\
23-03-2016 & 470.2059 & 0.265 & B & -0.938 & -0.957 \\
24-03-2016 & 471.2084 & 0.317 & V & -1.083 & -1.094 \\
24-03-2016 & 471.2084 & 0.317 & B & -0.978 & -0.991 \\
25-03-2016 & 472.2103 & 0.368 & V & -1.112 & -1.113 \\
25-03-2016 & 472.2103 & 0.368 & B & -1.002 & -1.003 \\
28-03-2016 & 475.2029 & 0.520 & V & -1.124 & -1.111 \\
28-03-2016 & 475.2029 & 0.520 & B & -1.020 & -1.006 \\
02-04-2016 & 480.2018 & 0.775 & V & -0.969 & -0.979 \\
10-04-2016 & 488.1953 & 0.183 & V & -1.025 & -1.036 \\
10-04-2016 & 488.1953 & 0.183 & B & -0.911 & -0.923 \\
16-04-2016 & 494.1596 & 0.487 & V & -1.127 & -1.113 \\
16-04-2016 & 494.1596 & 0.487 & B & -1.038 & -1.022 \\
17-04-2016 & 495.1729 & 0.539 & V & -1.124 & -1.113 \\
17-04-2016 & 495.1729 & 0.539 & B & -1.006 & -0.993 \\
18-04-2016 & 496.1647 & 0.590 & V & -1.094 & -1.089 \\
18-04-2016 & 496.1647 & 0.590 & B & -0.989 & -0.984 \\
21-04-2016 & 499.1678 & 0.743 & V & -0.978 & -0.989 \\
21-04-2016 & 499.1678 & 0.743 & B & -0.845 & -0.857 \\
26-04-2016 & 504.1629 & 0.998 & V & -0.981 & -0.967 \\
26-04-2016 & 504.1629 & 0.998 & B & -0.863 & -0.847 \\
27-04-2016 & 505.1652 & 0.049 & V & -0.983 & -0.970 \\
27-04-2016 & 505.1652 & 0.049 & B & -0.859 & -0.845 \\
29-04-2016 & 507.1679 & 0.151 & V & -1.021 & -1.026 \\
29-04-2016 & 507.1679 & 0.151 & B & -0.894 & -0.899 \\
30-04-2016 & 508.1606 & 0.202 & V & -1.051 & -1.065 \\
30-04-2016 & 508.1606 & 0.202 & B & -0.920 & -0.936 \\
01-05-2016 & 509.1611 & 0.253 & V & -1.060 & -1.077 \\
01-05-2016 & 509.1611 & 0.253 & B & -0.945 & -0.965 \\
03-05-2016 & 511.1612 & 0.355 & V & -1.114 & -1.118 \\
03-05-2016 & 511.1612 & 0.355 & B & -1.009 & -1.013 \\
05-05-2016 & 513.1497 & 0.456 & B & -1.026 & -1.011 \\
\label{obs_log_photom}
\end{longtable}
} 

\longtab[2]{
\begin{longtable}{cccccccc}
\caption{Spectroscopic observing log. The SNR is measured per pixel in a continuum region around 6550\,{\AA}.}\\
\hline
\hline
Date & Time & MJD & Orbital & Set & SNR & Combined & Combined\\
 &  & +57000 & Phase &  &   & phase & SNR\\
\hline
\endfirsthead
\caption{Continued.} \\
\hline
Date & Time & MJD & Orbital & Set & SNR  & Combined & Combined \\
 & UT & +57000 & Phase &  &   &  phase & SNR \\
\hline
\endhead
\hline
2015-11-04 & 01:48:22 &  330.076303 &  0.117 &  1 &  145  & & \\
2015-11-04 & 01:51:28 &  330.078455 &  0.117 &  1 &  159  & & \\
2015-11-05 & 03:36:19 &  331.151268 &  0.172 &  1 &  192  & & \\
2015-11-05 & 03:39:25 &  331.153421 &  0.172 &  1 &  241  & & \\
2015-11-06 & 02:49:52 &  332.119006 &  0.221 &  1 &  215  & & \\
2015-11-06 & 02:52:58 &  332.121159 &  0.221 &  1 &  192  & & \\
2015-11-07 & 02:42:01 &  333.113556 &  0.272 &  1 &  174  & & \\
2015-11-07 & 02:45:07 &  333.115709 &  0.272 &  1 &  162  & & \\
2015-11-12 & 02:29:45 &  338.105038 &  0.527 &  1 &  211  & & \\
2015-11-12 & 02:32:51 &  338.107191 &  0.527 &  1 &  200  & & \\
2015-11-14 & 02:05:43 &  340.088349 &  0.628 &  1 &  142  & & \\
2015-11-14 & 02:08:49 &  340.090502 &  0.628 &  1 &  121  & & \\
2015-11-18 & 00:22:32 &  344.016701 &  0.828 &  1 &  161  & & \\
2015-11-18 & 00:25:38 &  344.018854 &  0.828 &  1 &  146  & & \\
2015-11-20 & 03:09:50 &  346.132871 &  0.936 &  1 &  130  & & \\
2015-11-20 & 03:12:56 &  346.135023 &  0.936 &  1 &  159  & & \\
2015-11-21 & 03:19:59 &  347.139929 &  0.987 &  1 &  185  & & \\
2015-11-21 & 03:23:05 &  347.142082 &  0.988 &  1 &  168  & & \\
2015-11-22 & 03:17:53 &  348.138470 &  0.038 &  1 &  173  & & \\
2015-11-22 & 03:20:59 &  348.140624 &  0.038 &  1 &  197  & & \\
2015-11-24 & 03:10:57 &  350.133653 &  0.140 &  2 &  129  & & \\
2015-11-25 & 03:20:01 &  351.139947 &  0.191 &  2 &  185  & & \\
2015-11-25 & 03:23:07 &  351.142100 &  0.192 &  2 &  184  & & \\
2015-11-26 & 03:22:55 &  352.141960 &  0.243 &  2 &  202  & & \\
2015-11-26 & 03:26:01 &  352.144113 &  0.243 &  2 &  203  & & \\
2015-11-27 & 01:34:47 &  353.066869 &  0.290 &  2 &  169  & & \\
2015-11-27 & 01:37:53 &  353.069022 &  0.290 &  2 &  199  & & \\
2015-11-28 & 06:25:20 &  354.268640 &  0.351 &  2 &  176  & & \\
2015-11-28 & 06:28:26 &  354.270792 &  0.351 &  2 &  197  & & \\
2015-11-30 & 06:32:30 &  356.273614 &  0.453 &  2 &  184  & & \\
2015-11-30 & 06:35:36 &  356.275767 &  0.453 &  2 &  240  & & \\
2015-12-01 & 06:43:09 &  357.281018 &  0.505 &  2 &  156  & & \\
2015-12-01 & 06:46:15 &  357.283171 &  0.505 &  2 &  140  & & \\
2015-12-02 & 06:36:50 &  358.276622 &  0.556 &  2 &  144  & & \\
2015-12-02 & 06:39:56 &  358.278774 &  0.556 &  2 &  178  & & \\
2015-12-03 & 06:37:40 &  359.277202 &  0.607 &  2 &  163  & & \\
2015-12-03 & 06:40:46 &  359.279355 &  0.607 &  2 &  106  & & \\
2015-12-05 & 06:56:41 &  361.290409 &  0.709 &  2 &  134  & & \\
2015-12-10 & 06:34:19 &  366.274883 &  0.964 &  2 &   94  & & \\
2015-12-10 & 06:37:25 &  366.277035 &  0.964 &  2 &  107  & & \\
2015-12-11 & 07:13:17 &  367.301935 &  0.016 &  2 &  163  & & \\
2015-12-11 & 07:16:23 &  367.304088 &  0.016 &  2 &  146  & & \\
2015-12-12 & 06:48:07 &  368.284461 &  0.066 &  2 &  205  & & \\
2015-12-12 & 06:51:13 &  368.286614 &  0.066 &  2 &  160  & & \\
2015-12-13 & 07:22:31 &  369.308346 &  0.118 &  3 &  158  & & \\
2015-12-13 & 07:25:37 &  369.310499 &  0.118 &  3 &  250  & & \\
2015-12-14 & 06:01:17 &  370.251934 &  0.166 &  3 &  186  & & \\
2015-12-14 & 06:04:23 &  370.254086 &  0.167 &  3 &  222  & & \\
2015-12-15 & 06:20:53 &  371.265549 &  0.218 &  3 &  176  & & \\
2015-12-15 & 06:23:59 &  371.267702 &  0.218 &  3 &  185  & & \\
2015-12-17 & 06:37:10 &  373.276859 &  0.321 &  3 &  233  & & \\
2015-12-17 & 06:40:16 &  373.279012 &  0.321 &  3 &  380  & & \\
2015-12-18 & 05:47:41 &  374.242497 &  0.370 &  3 &  166  & & \\
2015-12-18 & 05:50:47 &  374.244650 &  0.370 &  3 &  179  & & \\
2015-12-19 & 05:43:15 &  375.239416 &  0.421 &  3 &  157  & & \\
2015-12-19 & 05:46:21 &  375.241569 &  0.421 &  3 &  176  & & \\
2015-12-20 & 03:45:00 &  376.157303 &  0.468 &  3 &  251  & & \\
2015-12-20 & 03:48:06 &  376.159456 &  0.468 &  3 &  247  & & \\
2015-12-22 & 05:31:35 &  378.231318 &  0.573 &  3 &  181  & & \\
2015-12-22 & 05:34:41 &  378.233472 &  0.574 &  3 &  177  & & \\
2015-12-23 & 06:27:38 &  379.270237 &  0.626 &  3 &  181  & & \\
2015-12-23 & 06:30:44 &  379.272391 &  0.627 &  3 &  162  & & \\
2015-12-24 & 05:19:51 &  380.223163 &  0.675 &  3 &  192  & & \\
2015-12-24 & 05:22:57 &  380.225317 &  0.675 &  3 &  277  & & \\
2015-12-28 & 05:04:01 &  384.212170 &  0.879 &  3 &  148  & & \\
2015-12-28 & 05:07:07 &  384.214323 &  0.879 &  3 &  227  & & \\
2015-12-29 & 05:00:41 &  385.209858 &  0.929 &  3 &  188  & & \\
2015-12-29 & 05:03:47 &  385.212010 &  0.930 &  3 &  131  & & \\
2015-12-30 & 04:56:09 &  386.206710 &  0.980 &  3 &  181  & & \\
2015-12-30 & 04:59:15 &  386.208862 &  0.980 &  3 &  150  & & \\
2015-12-31 & 04:53:32 &  387.204893 &  0.031 &  3 &  254  & & \\
2015-12-31 & 04:56:38 &  387.207045 &  0.031 &  3 &  178  & & \\
2016-01-04 & 03:02:33 &  391.127813 &  0.231 &  4 &  243  & 0.231 & 300\\
2016-01-04 & 03:05:39 &  391.129966 &  0.231 &  4 &  191  & 0.231 & 300\\
2016-01-05 & 03:23:47 &  392.142569 &  0.283 &  4 &  185  & & \\
2016-01-06 & 02:50:55 &  393.119743 &  0.333 &  4 &  210  & 0.333 & 249\\
2016-01-06 & 02:54:01 &  393.121896 &  0.333 &  4 &  174  & 0.333 & 249\\
2016-01-06 & 20:23:50 &  393.850932 &  0.370 &  4 &  170  & 0.370 & 222\\
2016-01-06 & 20:26:56 &  393.853087 &  0.370 &  4 &  211  & 0.370 & 222\\
2016-01-07 & 20:31:40 &  394.856375 &  0.422 &  4 &  233  & 0.422 & 258\\
2016-01-07 & 20:34:46 &  394.858528 &  0.422 &  4 &  226  & 0.422 & 258\\
2016-01-08 & 21:58:00 &  395.916327 &  0.476 &  4 &  206  & 0.476 & 309\\
2016-01-08 & 22:01:06 &  395.918480 &  0.476 &  4 &  238  & 0.476 & 309\\
2016-01-09 & 22:08:31 &  396.923624 &  0.527 &  4 &  224  & 0.527 & 278\\
2016-01-09 & 22:11:37 &  396.925777 &  0.527 &  4 &  199  & 0.527 & 278\\
2016-01-11 & 21:08:39 &  398.882060 &  0.627 &  4 &  206  & 0.627 & 207\\
2016-01-11 & 21:11:45 &  398.884213 &  0.627 &  4 &  186  & 0.627 & 207\\
2016-01-12 & 21:52:04 &  399.912207 &  0.679 &  4 &  212  & 0.680 & 258\\
2016-01-12 & 21:55:10 &  399.914361 &  0.680 &  4 &  197  & 0.680 & 258\\
2016-01-12 & 21:58:16 &  399.916512 &  0.680 &  4 &  187  & 0.680 & 258\\
2016-01-12 & 22:01:22 &  399.918663 &  0.680 &  4 &  174  & 0.680 & 258\\
2016-01-12 & 22:04:28 &  399.920814 &  0.680 &  4 &  180  & 0.680 & 258\\
2016-01-12 & 22:08:29 &  399.923610 &  0.680 &  4 &  168  & 0.680 & 258\\
2016-01-12 & 22:11:35 &  399.925763 &  0.680 &  4 &  193  & 0.680 & 258\\
2016-01-12 & 22:14:41 &  399.927914 &  0.680 &  4 &  184  & 0.680 & 258\\
2016-01-12 & 22:17:47 &  399.930065 &  0.680 &  4 &  193  & 0.680 & 258\\
2016-01-12 & 22:20:53 &  399.932218 &  0.681 &  4 &  180  & 0.680 & 258\\
2016-01-13 & 21:00:44 &  400.876555 &  0.729 &  4 &  182  & 0.730 & 255\\
2016-01-13 & 21:03:50 &  400.878710 &  0.729 &  4 &  189  & 0.730 & 255\\
2016-01-13 & 21:06:56 &  400.880860 &  0.729 &  4 &  113  & 0.730 & 255\\
2016-01-13 & 21:10:02 &  400.883011 &  0.729 &  4 &  154  & 0.730 & 255\\
2016-01-13 & 21:13:07 &  400.885161 &  0.729 &  4 &  151  & 0.730 & 255\\
2016-01-13 & 22:35:29 &  400.942353 &  0.732 &  4 &  184  & 0.730 & 255\\
2016-01-13 & 22:38:35 &  400.944507 &  0.732 &  4 &   93  & 0.730 & 255\\
2016-01-13 & 22:41:41 &  400.946659 &  0.732 &  4 &   55  & 0.730 & 255\\
2016-01-13 & 22:44:47 &  400.948809 &  0.732 &  4 &   64  & 0.730 & 255\\
2016-01-14 & 22:02:08 &  401.919195 &  0.782 &  4 &  176  & 0.782 & 303\\
2016-01-14 & 22:05:14 &  401.921349 &  0.782 &  4 &  133  & 0.782 & 303\\
2016-01-14 & 22:08:20 &  401.923500 &  0.782 &  4 &  162  & 0.782 & 303\\
2016-01-14 & 22:11:26 &  401.925651 &  0.782 &  4 &  173  & 0.782 & 303\\
2016-01-14 & 22:14:32 &  401.927801 &  0.782 &  4 &  167  & 0.782 & 303\\
2016-01-14 & 22:18:23 &  401.930483 &  0.782 &  4 &   95  & 0.782 & 303\\
2016-01-15 & 22:09:09 &  402.924063 &  0.833 &  4 &  126  & & \\
2016-01-17 & 00:01:06 &  404.001808 &  0.888 &  4 &  168  & 0.889 & 244\\
2016-01-17 & 00:04:12 &  404.003961 &  0.888 &  4 &  222  & 0.889 & 244\\
2016-01-17 & 00:07:17 &  404.006111 &  0.888 &  4 &  206  & 0.889 & 244\\
2016-01-17 & 00:10:23 &  404.008261 &  0.888 &  4 &  205  & 0.889 & 244\\
2016-01-17 & 00:13:29 &  404.010414 &  0.889 &  4 &  218  & 0.889 & 244\\
2016-01-17 & 00:17:54 &  404.013484 &  0.889 &  4 &  151  & 0.889 & 244\\
2016-01-17 & 00:21:01 &  404.015637 &  0.889 &  4 &  140  & 0.889 & 244\\
2016-01-17 & 00:24:06 &  404.017788 &  0.889 &  4 &  150  & 0.889 & 244\\
2016-01-17 & 00:27:12 &  404.019939 &  0.889 &  4 &  152  & 0.889 & 244\\
2016-01-17 & 00:30:18 &  404.022090 &  0.889 &  4 &  207  & 0.889 & 244\\
2016-01-18 & 20:00:05 &  405.834442 &  0.982 &  4 &  177  & 0.982 & 210\footnote{In this phase there is a small water line in the region where SNR is measured. This causes the SNR measurement, especially in the combined spectrum, to be underestimated.}\\

2016-01-18 & 20:03:11 &  405.836594 &  0.982 &  4 &  189  & 0.982 & 210$^{2}$\\
2016-01-18 & 20:06:17 &  405.838745 &  0.982 &  4 &  194  & 0.982 & 210$^{2}$\\
2016-01-18 & 20:09:23 &  405.840895 &  0.982 &  4 &  170  & 0.982 & 210$^{2}$\\
2016-01-18 & 20:12:29 &  405.843046 &  0.982 &  4 &  164  & 0.982 & 210$^{2}$\\
2016-01-18 & 20:16:36 &  405.845913 &  0.982 &  4 &  162  & 0.982 & 210$^{2}$\\
2016-01-18 & 20:19:42 &  405.848066 &  0.982 &  4 &  156  & 0.982 & 210$^{2}$\\
2016-01-18 & 20:22:48 &  405.850217 &  0.982 &  4 &  160  & 0.982 & 210$^{2}$\\
2016-01-18 & 20:25:54 &  405.852367 &  0.983 &  4 &  194  & 0.982 & 210$^{2}$\\
2016-01-19 & 20:47:21 &  406.867262 &  0.034 &  4 &  244  & 0.035 & 308\\
2016-01-19 & 20:50:27 &  406.869417 &  0.034 &  4 &  261  & 0.035 & 308\\
2016-01-19 & 20:53:33 &  406.871568 &  0.035 &  4 &  209  & 0.035 & 308\\
2016-01-19 & 20:56:39 &  406.873719 &  0.035 &  4 &  194  & 0.035 & 308\\
2016-01-19 & 20:59:45 &  406.875871 &  0.035 &  4 &  213  & 0.035 & 308\\
2016-01-20 & 20:24:27 &  407.851356 &  0.084 &  4 &  236  & 0.085 & 427\\
2016-01-20 & 20:27:33 &  407.853509 &  0.085 &  4 &  186  & 0.085 & 427\\
2016-01-20 & 20:30:38 &  407.855660 &  0.085 &  4 &  207  & 0.085 & 427\\
2016-01-20 & 20:33:44 &  407.857810 &  0.085 &  4 &  201  & 0.085 & 427\\
2016-01-20 & 20:36:50 &  407.859963 &  0.085 &  4 &  275  & 0.085 & 427\\
2016-01-20 & 20:40:49 &  407.862727 &  0.085 &  4 &  283  & 0.085 & 427\\
2016-01-20 & 20:43:55 &  407.864881 &  0.085 &  4 &  192  & 0.085 & 427\\
2016-01-20 & 20:47:01 &  407.867033 &  0.085 &  4 &  242  & 0.085 & 427\\
2016-01-20 & 20:50:07 &  407.869186 &  0.085 &  4 &  217  & 0.085 & 427\\
2016-01-20 & 20:53:13 &  407.871337 &  0.086 &  4 &  278  & 0.085 & 427\\
2016-01-21 & 20:58:11 &  408.874790 &  0.137 &  4 &  194  & 0.137 & 287\\
2016-01-21 & 21:01:17 &  408.876944 &  0.137 &  4 &  262  & 0.137 & 287\\
2016-01-21 & 21:04:23 &  408.879095 &  0.137 &  4 &  253  & 0.137 & 287\\
2016-01-23 & 19:34:29 &  410.816660 &  0.236 &  5 &  175  & 0.236 & 310\\
2016-01-23 & 19:37:35 &  410.818814 &  0.236 &  5 &  159  & 0.236 & 310\\
2016-01-23 & 19:40:41 &  410.820965 &  0.236 &  5 &  191  & 0.236 & 310\\
2016-01-23 & 19:43:47 &  410.823117 &  0.236 &  5 &  181  & 0.236 & 310\\
2016-01-23 & 19:46:53 &  410.825268 &  0.236 &  5 &  165  & 0.236 & 310\\
2016-01-23 & 19:50:44 &  410.827951 &  0.236 &  5 &  189  & 0.236 & 310\\
2016-01-23 & 19:53:50 &  410.830104 &  0.236 &  5 &  162  & 0.236 & 310\\
2016-01-23 & 19:56:56 &  410.832255 &  0.237 &  5 &  153  & 0.236 & 310\\
2016-01-23 & 20:00:02 &  410.834406 &  0.237 &  5 &  159  & 0.236 & 310\\
2016-01-23 & 20:03:08 &  410.836556 &  0.237 &  5 &  163  & 0.236 & 310\\
2016-01-24 & 20:00:52 &  411.834987 &  0.288 &  5 &  141  & 0.288 & 152\\
2016-01-24 & 20:03:58 &  411.837139 &  0.288 &  5 &   75  & 0.288 & 152\\
2016-01-25 & 20:14:41 &  412.844582 &  0.339 &  5 &  213  & 0.339 & 295\\
2016-01-25 & 20:17:47 &  412.846734 &  0.339 &  5 &  174  & 0.339 & 295\\
2016-01-25 & 20:20:53 &  412.848885 &  0.339 &  5 &  180  & 0.339 & 295\\
2016-01-25 & 20:23:59 &  412.851035 &  0.340 &  5 &  214  & 0.339 & 295\\
2016-01-26 & 20:16:59 &  413.846170 &  0.390 &  5 &  195  & 0.391 & 307\\
2016-01-26 & 20:20:05 &  413.848322 &  0.390 &  5 &  229  & 0.391 & 307\\
2016-01-26 & 20:23:10 &  413.850473 &  0.391 &  5 &  275  & 0.391 & 307\\
2016-01-26 & 20:26:16 &  413.852623 &  0.391 &  5 &  170  & 0.391 & 307\\
2016-01-26 & 20:29:22 &  413.854774 &  0.391 &  5 &  197  & 0.391 & 307\\
2016-01-26 & 20:33:21 &  413.857536 &  0.391 &  5 &  229  & 0.391 & 307\\
2016-01-26 & 20:36:27 &  413.859689 &  0.391 &  5 &  213  & 0.391 & 307\\
2016-01-26 & 20:39:33 &  413.861841 &  0.391 &  5 &  191  & 0.391 & 307\\
2016-01-26 & 20:42:38 &  413.863993 &  0.391 &  5 &  198  & 0.391 & 307\\
2016-01-26 & 20:45:44 &  413.866144 &  0.391 &  5 &  205  & 0.391 & 307\\
2016-01-27 & 20:17:07 &  414.846269 &  0.441 &  5 &  217  & 0.442 & 334\\
2016-01-27 & 20:20:13 &  414.848426 &  0.441 &  5 &  239  & 0.442 & 334\\
2016-01-27 & 20:23:19 &  414.850577 &  0.442 &  5 &  281  & 0.442 & 334\\
2016-01-27 & 20:26:25 &  414.852728 &  0.442 &  5 &  213  & 0.442 & 334\\
2016-01-27 & 20:29:31 &  414.854879 &  0.442 &  5 &  224  & 0.442 & 334\\
2016-01-27 & 20:33:24 &  414.857577 &  0.442 &  5 &  212  & 0.442 & 334\\
2016-01-27 & 20:36:30 &  414.859732 &  0.442 &  5 &  274  & 0.442 & 334\\
2016-01-27 & 20:39:36 &  414.861883 &  0.442 &  5 &  186  & 0.442 & 334\\
2016-01-27 & 20:42:42 &  414.864035 &  0.442 &  5 &  234  & 0.442 & 334\\
2016-01-27 & 20:45:48 &  414.866189 &  0.442 &  5 &  185  & 0.442 & 334\\
2016-01-30 & 21:13:53 &  417.885683 &  0.596 &  5 &  124  & & \\
2016-02-01 & 20:23:07 &  419.850431 &  0.697 &  5 &  168  & & \\
2016-02-03 & 20:01:04 &  421.835124 &  0.798 &  5 &  186  & 0.798 & 338\\
2016-02-03 & 20:04:10 &  421.837277 &  0.798 &  5 &  207  & 0.798 & 338\\
2016-02-04 & 20:19:02 &  422.847604 &  0.849 &  5 &  101  & 0.850 & 235\\
2016-02-04 & 20:22:09 &  422.849757 &  0.850 &  5 &  164  & 0.850 & 235\\
2016-02-04 & 20:25:14 &  422.851908 &  0.850 &  5 &  149  & 0.850 & 235\\
2016-02-05 & 20:03:00 &  423.836467 &  0.900 &  5 &  172  & 0.900 & 264\\
2016-02-05 & 20:06:07 &  423.838624 &  0.900 &  5 &  160  & 0.900 & 264\\
2016-02-05 & 20:09:12 &  423.840775 &  0.900 &  5 &  165  & 0.900 & 264\\
2016-02-05 & 20:12:18 &  423.842926 &  0.900 &  5 &  173  & 0.900 & 264\\
2016-02-05 & 20:15:24 &  423.845078 &  0.900 &  5 &  195  & 0.900 & 264\\
2016-02-07 & 20:05:35 &  425.838261 &  0.002 &  5 &  128  & 0.003 & 251\\
2016-02-07 & 20:08:41 &  425.840414 &  0.002 &  5 &  176  & 0.003 & 251\\
2016-02-07 & 20:11:47 &  425.842566 &  0.002 &  5 &  235  & 0.003 & 251\\
2016-02-07 & 20:14:53 &  425.844717 &  0.002 &  5 &  198  & 0.003 & 251\\
2016-02-07 & 20:17:59 &  425.846867 &  0.002 &  5 &  174  & 0.003 & 251\\
2016-02-07 & 20:22:14 &  425.849818 &  0.003 &  5 &  216  & 0.003 & 251\\
2016-02-07 & 20:25:20 &  425.851971 &  0.003 &  5 &  127  & 0.003 & 251\\
2016-02-07 & 20:28:26 &  425.854122 &  0.003 &  5 &  108  & 0.003 & 251\\
2016-02-07 & 20:31:32 &  425.856274 &  0.003 &  5 &   77  & 0.003 & 251\\
2016-02-09 & 20:05:25 &  427.838138 &  0.104 &  5\&6 &  260  & 0.105 & 331\\
2016-02-09 & 20:08:31 &  427.840291 &  0.104 &  5\&6 &  280  & 0.105 & 331\\
2016-02-09 & 20:11:36 &  427.842441 &  0.104 &  5\&6 &  228  & 0.105 & 331\\
2016-02-09 & 20:14:42 &  427.844592 &  0.104 &  5\&6 &  192  & 0.105 & 331\\
2016-02-09 & 20:17:48 &  427.846743 &  0.105 &  5\&6 &  185  & 0.105 & 331\\
2016-02-09 & 20:22:01 &  427.849675 &  0.105 &  5\&6 &  189  & 0.105 & 331\\
2016-02-09 & 20:25:07 &  427.851829 &  0.105 &  5\&6 &  194  & 0.105 & 331\\
2016-02-09 & 20:28:13 &  427.853980 &  0.105 &  5\&6 &  167  & 0.105 & 331\\
2016-02-09 & 20:31:19 &  427.856130 &  0.105 &  5\&6 &  179  & 0.105 & 331\\
2016-02-09 & 20:34:25 &  427.858281 &  0.105 &  5\&6 &  185  & 0.105 & 331\\
2016-02-11 & 19:55:55 &  429.831540 &  0.206 &  6 &  176  & & \\
2016-02-11 & 19:59:01 &  429.833694 &  0.206 &  6 &  183  & & \\
2016-02-11 & 20:02:06 &  429.835845 &  0.206 &  6 &  152  & & \\
2016-02-11 & 20:05:12 &  429.837995 &  0.206 &  6 &  190  & & \\
2016-02-11 & 20:08:18 &  429.840146 &  0.206 &  6 &  147  & & \\
2016-02-11 & 20:12:10 &  429.842830 &  0.206 &  6 &  194  & & \\
2016-02-11 & 20:15:16 &  429.844983 &  0.206 &  6 &  187  & & \\
2016-02-11 & 20:18:22 &  429.847136 &  0.207 &  6 &  167  & & \\
2016-02-11 & 20:21:28 &  429.849286 &  0.207 &  6 &  249  & & \\
2016-02-11 & 20:24:34 &  429.851437 &  0.207 &  6 &  170  & & \\
2016-02-12 & 19:41:01 &  430.821204 &  0.256 &  6 &  171  & & \\
2016-02-12 & 19:44:07 &  430.823357 &  0.256 &  6 &  219  & & \\
2016-02-14 & 19:56:35 &  432.832003 &  0.359 &  6 &  220  & & \\
2016-02-14 & 19:59:41 &  432.834156 &  0.359 &  6 &  220  & & \\
2016-02-14 & 20:02:46 &  432.836308 &  0.359 &  6 &  202  & & \\
2016-02-14 & 20:05:52 &  432.838459 &  0.359 &  6 &  246  & & \\
2016-02-16 & 19:32:57 &  434.815601 &  0.460 &  6 &  166  & & \\
2016-02-17 & 19:36:36 &  435.818134 &  0.511 &  6 &  223  & & \\
2016-02-17 & 19:39:42 &  435.820288 &  0.511 &  6 &  226  & & \\
2016-02-17 & 19:42:48 &  435.822438 &  0.511 &  6 &  206  & & \\
2016-02-17 & 19:45:54 &  435.824590 &  0.511 &  6 &  232  & & \\
2016-02-26 & 03:16:51 &  444.137751 &  0.936 &  6 &  173  & & \\
2016-02-26 & 03:19:57 &  444.139904 &  0.936 &  6 &  183  & & \\
2016-02-26 & 03:23:03 &  444.142055 &  0.936 &  6 &  134  & & \\
2016-03-01 & 03:23:04 &  448.142067 &  0.140 &  7 &  272  & & \\
2016-03-04 & 03:00:55 &  451.126687 &  0.292 &  7 &  122  & & \\
2016-03-04 & 03:04:01 &  451.128840 &  0.292 &  7 &  103  & & \\
2016-03-04 & 03:07:07 &  451.130990 &  0.292 &  7 &  153  & & \\
2016-03-04 & 03:10:13 &  451.133141 &  0.292 &  7 &  129  & & \\
2016-03-06 & 19:58:32 &  453.833358 &  0.430 &  7 &  243  & & \\
2016-03-06 & 20:01:38 &  453.835511 &  0.430 &  7 &  212  & & \\
2016-03-06 & 20:04:44 &  453.837662 &  0.430 &  7 &  230  & & \\
2016-03-10 & 19:38:11 &  457.819229 &  0.633 &  7 &  191  & & \\
2016-03-10 & 19:41:17 &  457.821384 &  0.634 &  7 &  204  & & \\
2016-03-11 & 20:31:03 &  458.855946 &  0.686 &  7 &  163  & & \\
2016-03-11 & 20:34:09 &  458.858098 &  0.686 &  7 &  184  & & \\
2016-03-13 & 19:39:50 &  460.820379 &  0.787 &  7 &  202  & & \\
2016-03-13 & 19:42:56 &  460.822532 &  0.787 &  7 &  259  & & \\
2016-03-17 & 19:42:07 &  464.821957 &  0.991 &  7 &  185  & & \\
2016-03-17 & 19:45:13 &  464.824111 &  0.991 &  7 &  173  & & \\
2016-03-19 & 19:50:51 &  466.828024 &  0.093 &  7 &  145  & & \\
2016-03-19 & 19:53:57 &  466.830177 &  0.093 &  7 &  188  & & \\
2016-03-24 & 19:45:59 &  471.824645 &  0.348 &  7 &  179  & & \\
2016-03-24 & 19:49:05 &  471.826799 &  0.348 &  7 &  200  & & \\
2016-03-26 & 19:47:23 &  473.825620 &  0.450 &  7 &  197  & & \\
2016-03-26 & 19:50:29 &  473.827774 &  0.450 &  7 &  171  & & \\
2016-04-01 & 21:51:26 &  479.911760 &  0.760 &  7 &  198  & & \\
2016-04-01 & 21:54:32 &  479.913913 &  0.761 &  7 &  213  & & \\
\label{obs_log_spec}
\end{longtable}
}

\end{appendix}

\end{document}